\documentclass[12pt]{article}
\usepackage{amssymb,amsmath,epsfig}
\usepackage{graphicx}
\allowdisplaybreaks

\begin{document}
\title{\bf Physical Analysis of Spherical Stellar Structures in $f(\mathrm{Q},\mathrm{T})$ Theory}
\author{M. Zeeshan Gul \thanks{mzeeshangul.math@gmail.com}~, M. Sharif \thanks {msharif.math@pu.edu.pk}
~and Adeeba Arooj \thanks{aarooj933@gmail.com}\\
Department of Mathematics and Statistics, The University of Lahore,\\
1-KM Defence Road Lahore-54000, Pakistan.}
\date{}
\maketitle

\begin{abstract}
This paper explores the viability and stability of compact stellar
objects characterized by anisotropic matter in the framework of
$f(\mathrm{Q},\mathrm{T})$ theory, where $\mathrm{Q}$ denotes
non-metricity and $\mathrm{T}$ represents the trace of the
energy-momentum tensor. We consider a specific model of this theory
to obtain explicit expressions for the field equations governing the
behavior of matter and geometry in this context. Furthermore, the
Karmarkar condition is employed to assess the configuration of
static spherically symmetric structures. The values of unknown
constants in the metric potentials are determined through matching
conditions of the interior and exterior spacetimes. Various physical
quantities such as fluid parameters, energy constraints, equation of
state parameters, mass, compactness and redshift are graphically
analyzed to evaluate the viability of the considered compact stars.
The Tolman-Oppenheimer-Volkoff equation is used to examine the
equilibrium state of the stellar models. Moreover, the stability of
the proposed compact stars is investigated through sound speed and
adiabatic index methods. This study concludes that the proposed
compact stars analyzed in this theoretical framework are viable and
stable, as all the required conditions are satisfied.
\end{abstract}

\section{Introduction}

Stars serve as vital constituents of galaxies and maintain
equilibrium by balancing the inward gravitational force with the
outward pressure generated through nuclear fusion reactions. When a
star depletes its nuclear fuel, it may no longer generate sufficient
pressure to counteract gravitational collapse. This results in the
creation of compact objects such as white dwarfs, neutron stars or
black holes, depending on their initial mass. Neutron stars are the
most captivating cosmic objects that play pivotal roles in various
astrophysical processes. The intense magnetic fields surrounding
neutron stars lead to the emission of intense radiation,
particularly in the form of X-rays and gamma rays. Furthermore,
neutron stars have significant influence in astrophysical phenomena,
including the formation of heavy elements in the universe and the
generation of gravitational waves.

Advances in observational techniques, such as the development of
X-rays and radio telescopes have enabled scientists to make
significant progress in the detection and study of neutron stars.
These observations have not only opened up novel avenues for
research but have also deepened our comprehension of the fundamental
processes that govern the universe. Baade and Zwicky \cite{1} argued
that compact stars (CSs) are formed as a result of supernovae and
their existence has been proved by the discovery of pulsars
\cite{2}. Pulsars have captured the attention of many researchers
due to their extraordinary attributes such as extreme density,
powerful magnetic fields and rapid rotation, which results in
neutron stars emitting beams of electromagnetic radiation. These
beams manifest as regular pulses of radiation, leading to the name
``pulsar''. The exploration of neutron stars and pulsars provides
researchers with insights into diverse aspects of these captivating
objects, including their mass, radius, magnetic fields, rotational
dynamics and behavior under extreme conditions. The viable behavior
of pulsars with various consideration has been studied in \cite{3}.
Mak and Harko \cite{4} discussed viable features of rotating neutron
stars and their stability through the Herrera cracking approach.

Researchers have employed various approaches, including solving
metric elements, imposing constraints on fluid parameters and
utilizing specific equations of state to study the geometry of CSs.
Among these methods, the Karmarkar method stands out as a valuable
mathematical tool for exploring the interior geometry of CSs. This
method has been widely applied by numerous researchers to gain
insights into the physical characteristics and stability of the
cosmic objects. By establishing a correlation between temporal and
radial elements, the Karmarkar method aids in comprehending the
nature of the celestial objects. Specifically, the Karmarkar
condition is formulated for analyzing solutions in the context of
static spherical spacetime \cite{5}. Maurya et al \cite{6} delved
into the physically viable characteristics of anisotropic CSs using
this technique. Singh and Pant \cite{7} used the Karmarkar condition
to investigate the interior geometry of celestial objects. Bhar et
al \cite{8} employed this method to analyze the viability and
stability of anisotropic CSs.

Einstein's general theory of relativity (EGTR) is a fundamental
theory in physics that has revolutionized our comprehension of
gravity and the nature of spacetime. Regarded as the cornerstone of
modern physics, EGTR relies on Riemannian geometry which involves
the mathematical notion of a metric space. In this framework,
spacetime is characterized by geometric structures defined by
Riemann's metric. Additionally, EGTR's expansion provides insights
into the gravitational field and the behavior of matter on cosmic
scales. In contrast, Weyl \cite{9} developed a more comprehensive
geometric framework by introducing the concept of the length
connection, aiming to unify the fundamental forces of nature. This
connection does not primarily concern with the length of vectors
during parallel transport but rather focuses on adjusting or gauging
the conformal factor. Weyl's theory introduces the notion of
non-metricity by positing that the covariant divergence of the
metric tensor is non-zero.

Non-Riemannian geometries serve as extensions to Riemannian
geometry, offering broader descriptions of spacetime curvature
through the inclusion of torsion and non-metricity concepts. The
non-metricity scalar is a mathematical quantity which provides an
alternative explanation for cosmic expansion beyond the standard
cosmological model, which relies on dark energy \cite{9a}.
Teleparallel gravity is introduced as an alternative theory to EGTR,
where torsion ($\mathcal{T}$) represents gravitational interaction.
Unlike the Levi-Civita connection in EGTR, the teleparallel
equivalent to EGTR exhibits both non-metricity and zero curvature.
The integral actions for torsion curvature and non-metricity are
formulated as $\int\sqrt{-\mathrm{g}} \mathcal{T}$ \cite{10} and
$\int\sqrt{-\mathrm{g}} \mathrm{Q}$ \cite{11}, respectively.
Researchers have developed a range of extended gravitational
theories to address these challenges and delve deeper into the
mysteries of the cosmos \cite{11a}-\cite{11e}.

The modified symmetric teleparallel theory is further extended by
incorporating the trace of stress-energy tensor in the functional
action, named as $f(\mathrm{Q},\mathrm{T})$ theory \cite{12}. This
extension establishes a specific coupling between the trace of the
energy-momentum tensor and non-metricity, attracting significant
attention of researchers for its profound implications in the realm
of gravitational physics. The modifications introduced by
$f(\mathrm{Q}, \mathrm{T})$ gravity have an impact on the internal
structure of CSs. This influences changes in the relationship
between pressure and density, variations in stellar radii and mass
profiles. The corresponding equations of motion play a role in
hydrostatic equilibrium and affect the stability of the star.
Deviations from the predictions of EGTR may rise to unique
mass-radius relations that can be tested against observational data
from X-ray binaries. Neutron stars are prime sources of
gravitational waves in binary systems. The modifications introduced
by $f(\mathrm{Q}, \mathrm{T})$ gravity can lead to distinct
gravitational wave signatures that may differ from those predicted
by EGTR. Furthermore, the implications of $f(\mathrm{Q},
\mathrm{T})$ gravity may extend to other properties like surface
redshift, providing avenues for distinguishing this gravity model
from other theories in the context of CSs.

The presence of additional correction terms in this modified
framework has been underscored by noteworthy outcomes, potentially
exerting a substantial influence on the geometric viability in
comparison to EGTR. The $f(\mathrm{Q},\mathrm{T})$ theory is
employed as a mathematical tool to scrutinize obscure aspects of
gravitational dynamics on a large scale. Investigations into this
modified theory have extended to its implications for cosmic models
and the accelerated expansion of the universe. The motivation
driving the exploration of this theory encompasses an examination of
its theoretical implications, its alignment with observational data
and its significance in cosmological contexts. Arora et al \cite{13}
assured that $f(\mathrm{Q},\mathrm{T})$ gravity discusses late-time
acceleration of the cosmos. Bhattacharjee et al \cite{14}
scrutinized the phenomenon of baryogenesis in the
$f(\mathrm{Q},\mathrm{T})$ theory. Arora and Sahoo \cite{14a}
studied the transitions between accelerated and decelerated phases
of the universe expansion through deceleration parameter, indicating
that the dynamics of cosmic expansion can be accounted for
$f(\mathrm{Q},\mathrm{T})$ theory. Xu et al \cite{14b} predicted a
de Sitter type cosmic expansion in this framework proposing it as an
alternative explanation of dark energy. Najera and Fajardo
\cite{14c} assured that this gravity offers alternation to the
standard cosmological model ($\Lambda$CDM) suggesting deviations
from conventional frameworks. Godani and Samanta \cite{14d} studied
the cosmic evolution using various cosmic parameters in this theory,
as it accounts for the accelerated cosmic expansion. Agrawal et al
\cite{14e} demonstrated the existence of matter bounce scenario in
$f(\mathrm{Q},\mathrm{T})$ gravity, which offers alternative
cosmological scenario to the Big Bang. Pradhan et al \cite{14f}
studied physical properties to ensure the existence of a stable
gravastar model within the modified theory that suggest alternative
configurations for CSs.

The intriguing properties of CSs yield captivating results in the
context of alternative gravitational theories. Arapoglu et al
\cite{16} employed perturbation techniques to investigate the
geometry of CSs in $f(\mathrm{R})$ gravity. They explored how
deviations from the EGTR impact the structure of compact objects. In
the same theoretical framework, Astashenok et al \cite{17} delved
into the structural aspects of pulsars with anisotropic matter
configurations. Das et al \cite{18} explored the influence of
effective matter variables on the geometry of anisotropic spheres in
$f(\mathrm{R},\mathrm{T})$ theory. Deb et al \cite{19} analyzed the
geometry of spherically symmetric strange stars in a similar
context. Biswas et al \cite{20} discussed strange quark stars using
the Krori-Barua solution in the context of curvature-matter coupled
theory. Bhar et al \cite{21} explored the viable characteristics of
a 4U 1538-52 CS in Einstein Gauss-Bonnet gravity. In $f(\mathrm{G})$
theory, Sharif and Ramzan \cite{22} studied the behavior of various
physical quantities and the stability of distinct CSs. Dey et al
\cite{23} employed the Finch-Skea symmetry to examine the viability
of anisotropic stellar models in $f(\mathrm{R},\mathrm{T})$ theory.
Sharif and his collaborators studied the studied the Noether
symmetry approach \cite{23a}-\cite{00000023a}, stability of the
Einstein universe \cite{23b}-\cite{0023b} and dynamics of
gravitational collapse \cite{23c}-\cite{000023c} and static
spherically symmetric structures \cite{23d}-\cite{0023d} in
$f(\mathrm{R},\mathrm{T}^{2})$ theory. Recently, the study of
observational constraints in modified gravities discussed in
\cite{24a}-\cite{24b}.

The above literature motivates us to investigate the viable
characteristics of anisotropic CSs in the context of
$f(\mathrm{Q},\mathrm{T})$ theory. We use the following format in
the paper. Section \textbf{2} presents the fundamental formulation
of $f(\mathrm{Q},\mathrm{T})$ gravity. In section \textbf{3}, a
specific model of this theory is considered to derive explicit
expressions for energy density and pressure components. We evaluate
unknown parameters through the matching conditions. In Section
\textbf{4}, various physical quantities are used to determine the
physical features of the considered CSs. Furthermore, the
equilibrium state and the stability of the considered CSs are
analyzed in Section \textbf{5}. We compile our findings in Section
\textbf{6}

\section{$f(\mathrm{Q},\mathrm{T})$ Theory: Basic Formalism}

This section introduces the fundamental framework of the modified
$f(\mathrm{Q},\mathrm{T})$ theory and outlines the derivation of
field equations through the variational principle. Weyl \cite{9}
extended Riemannian geometry to serve as a mathematical foundation
for describing gravitation in the framework of EGTR. In Riemannian
geometry, the direction and length of a vector remain unchanged
during parallel transport around a closed path. However, Weyl
proposed a modification where a vector undergoes both directional
and length changes during such transport. This modification
introduces a new vector field $(h^{\mu})$ that characterizes the
geometric aspects of Weyl geometry. The key fields in Weyl's space
include this new vector field and the metric tensor. While the
metric tensor defines the local structure of spacetime, specifying
distances and angles, the vector field is introduced to account for
length changes during parallel transport. Weyl's theory posits that
the vector field shares mathematical properties with electromagnetic
potentials in physics, suggesting a profound connection between
gravitational and electromagnetic forces. Weyl's proposal implies a
common geometric origin for these forces \cite{25}.

In Weyl geometry, transporting a vector along an infinitesimal path
results in a change in its length as $\delta l= lh_{\mu} \delta x^
{\mu}$ \cite{25}. This signifies that the variation in the vector's
length is directly proportional to the original length the
connection coefficient and the displacement along the path. When the
vector is transported in parallel around a small closed loop in Weyl
space, the variation in its length can be described as follows
\begin{eqnarray}\label{1}
\delta l= l\Psi_{\mu\nu}\delta s^{\mu\nu}, \quad
\Psi_{\mu\nu}=\nabla_{\nu}h_{\mu}-\nabla_{\mu}h_{\nu}.
\end{eqnarray}
This states that the variation in the vector's length is
proportional to the original length $(l)$, the curvature of the Weyl
connection $(\Psi)$ and the area enclosed by the loop $(\delta
s^{\mu\nu})$. A local scaling length of the form $\bar{l}=\phi(x)l$
changes the field $h_{\mu}$ to $\bar{h_{\mu}}=h_{\mu}+
(\ln\phi),_{\mu}$, whereas the elements of metric tensor are
modified by the conformal transformations
$\bar{\mathrm{g}}_{\mu\nu}=\phi^{2}\mathrm{g}_{\mu\nu}$ and
$\bar{\mathrm{g}}^{\mu\nu}=\phi^{-2}\mathrm{g}^{\mu\nu}$,
respectively \cite{26}. A semi-metric connection is another
important feature of the Weyl geometry, defined as
\begin{equation}\label{2}
\bar{\Gamma}^{\lambda}_{\mu\nu}=\Gamma^{\lambda}_{\mu\nu}
+\mathrm{g}_{\mu\nu}h^{\lambda}-\delta^{\lambda}_{\mu}h_{\nu}-
\delta^{\lambda}_{\nu}h_{\mu},
\end{equation}
where $\Gamma^{\lambda}_{\mu\nu}$ denotes the Christoffel symbol.
One can construct a gauge covariant derivative based on the
supposition that $ \bar{\Gamma}^{\lambda}_{\mu\nu}$ is symmetric
\cite{26}. The Weyl curvature tensor using the covariant derivative
can be expressed as
\begin{equation}\label{3}
\bar{\mathcal{C}}_{\mu\nu\lambda\xi}=\bar{\mathcal{C}}
_{(\mu\nu)\lambda\xi}+\bar{\mathcal{C}}_{[\mu\nu] \lambda\xi},
\end{equation}
where
\begin{eqnarray}\nonumber
\bar{\mathcal{C}}_{[\mu\nu]\lambda\xi}&=&\mathcal{C}
_{\mu\nu\lambda\xi}+2\nabla_{\lambda}h_{[\mu
\mathrm{g}_{\nu}]\xi}+2\nabla_{\xi}h_{[\nu\mathrm{g}_{\mu}]\lambda}
+2h_{\lambda}h_{[\mu\mathrm{g}_{\nu}]\xi}+2h_{\xi}h_{[\nu
\mathrm{g}_{\mu}]\lambda}
\\\nonumber
&-&2h^{2}\mathrm{g}_{\lambda[\mu\mathrm{g}_{\nu}]\xi},
\\\nonumber
\bar{\mathcal{C}}_{(\mu\nu)\lambda\xi}&=&
\frac{1}{2}(\bar{\mathcal{C}}_{\mu\nu\lambda\xi}
+\bar{\mathcal{C}}_{\nu\mu\lambda\xi}),
\\\nonumber
&=&\mathrm{g}_{\mu\nu}\Psi_{\lambda\xi}.
\end{eqnarray}
The Weyl curvature tensor after the first contraction yields
\begin{equation}\label{4}
\bar{\mathcal{C}}^{\mu}_{\nu}=\bar{\mathcal{C}}
^{\lambda\mu}_{\lambda\nu}=\mathcal{C}^{\mu}_{\nu}
+2h^{\mu}h_{\nu}+3\nabla_{\nu}h^{\mu}-\nabla_{\mu}h^{\nu}
+g^{\mu}_{\nu}(\nabla_{\lambda}h^{\lambda}-2h_{\lambda}h^{\lambda}).
\end{equation}
Finally, we obtain Weyl scalar as
\begin{equation}\label{5}
\bar{\mathcal{C}}=\bar{\mathcal{C}}^{\lambda}_{\lambda}=
\mathcal{C}+6(\nabla_{\mu}h^{\mu}-h_{\mu}h^{\mu}).
\end{equation}

Weyl-Cartan (WC) spaces incorporates torsion which provides a more
expansive framework compared to Riemannian and Weyl geometry. This
extended geometric structure serves as a versatile model for gravity
theories, accommodating diverse scales and behaviors in parallel
transport. Such investigations are particularly relevant for
exploring alternative gravity theories or pursuing the unification
of gravity with other fundamental forces. In a WC spacetime, vector
length is defined by a symmetric metric tensor, and the rule for
parallel transport is dictated by an asymmetric connection expressed
as $d\varpi^{\mu} =
-\varpi^{\lambda}{\Gamma}^{\mu}_{\lambda\nu}dx^{\nu}$ \cite{27}. The
connection for the WC geometry is expressed as
\begin{equation}\label{6}
\tilde{\Gamma}^{\lambda}_{\mu\nu}={\Gamma}^{\lambda}_{\mu\nu}
+\mathcal{W}^{\lambda}_{\mu}+\mathrm{L}^{\lambda}_{\mu\nu},
\end{equation}
where $ \mathcal{W}^{\lambda}_{\mu\nu}$ is the contortion tensor and
$ \mathrm{L}^{\lambda}_{\mu\nu}$ is the deformation tensor. The
contorsion tensor from the torsion tensor can be obtained as
\begin{equation}\label{7}
\mathcal{W}^{\lambda}_{\mu\nu}=\tilde{\Gamma}^{\lambda}_{[\mu\nu]}
+\mathrm{g}^{\lambda\xi}\mathrm{g}_{\mu\varsigma}
\tilde{\Gamma}^{\varsigma}_{[\nu\xi]}+\mathrm{g}^{\lambda\xi}
\mathrm{g}_{\nu\varsigma} \tilde{\Gamma}^{\varsigma}_{[\mu\xi]}.
\end{equation}
The non-metricity yields the deformation tensor as
\begin{equation}\label{8}
\mathrm{L}^{\lambda}_{\mu\nu}=\frac{1}{2}\mathrm{g}^{\lambda\xi}
(\mathrm{Q}_{\nu\mu\xi}
+\mathrm{Q}_{\mu\nu\xi}-\mathrm{Q}_{\lambda\mu\nu}),
\end{equation}
where
\begin{equation}\label{9}
\mathrm{Q}_{\lambda\mu\nu}=\nabla_{\lambda} \mathrm{g}_{\mu\nu}
=-\partial\mathrm{g}_{\mu\nu,\lambda}+\mathrm{g}_{\nu\xi}
\tilde{\Gamma}^{\xi}_{\mu\lambda}
+\mathrm{g}_{\xi\mu}\tilde{\Gamma}^{\xi}_{\nu\lambda},
\end{equation}
and $\tilde{\Gamma}^{\lambda}_{\mu\nu}$ is WC connection. Equations
(\ref{2}) and (\ref{6}) indicate that the WC geometry with zero
torsion is a particular case of Weyl geometry, where the
non-metricity is defined as
$\mathrm{Q}_{\lambda\mu\nu}=-2\mathrm{g}_{\mu\nu}h_{\lambda}$.
Therefore, Eq.(\ref{6}) turns out to be
\begin{equation}\label{10}
\tilde{\Gamma}^{\lambda}_{\mu\nu}={\Gamma}^{\lambda}_{\mu\nu}
+\mathrm{g}_{\mu\nu}h^{\lambda}
-\delta^{\lambda}_{\mu}h_{\nu}-\delta^{\lambda}_{\nu}h_{\mu}
+\mathcal{W}^{\lambda}_{\mu\nu},
\end{equation}
where
\begin{equation}\label{11}
\mathcal{W}^{\lambda}_{\mu\nu}=\mathcal{T}^{\lambda}
_{\mu\nu}-\mathrm{g}^{\lambda\xi}\mathrm{g}_{\varsigma\mu}
\mathcal{T}^{\varsigma}_{\xi\nu}-\mathrm{g}^{\lambda\xi}
\mathrm{g}_{\varsigma\nu}
\mathcal{T}^{\varsigma}_{\xi\mu},
\end{equation}
is the contortion and the WC torsion is expressed as
\begin{equation}\label{12}
\mathcal{T}^{\lambda}_{\mu\nu}=\frac{1}{2} (\tilde{\Gamma}^{\lambda}
_{\mu\nu}-\tilde{\Gamma}^{\lambda}_{\nu\mu}).
\end{equation}
The WC curvature tensor with the use of the connection is defined as
\begin{equation}\label{13}
\tilde{\mathcal{C}}^{\lambda}_{\mu\nu\xi} =\tilde{\Gamma}^{\lambda}
_{\mu\xi,\nu} -\tilde{\Gamma}^{\lambda}_{\mu\nu,\xi}+\tilde{\Gamma}
^{\lambda}_{\mu\xi}
\tilde{\Gamma}^{\varsigma}_{\lambda\nu}-\tilde{\Gamma}
^{\lambda}_{\mu\nu} \tilde{\Gamma}^{\varsigma}_{\lambda\xi}.
\end{equation}
The WC scalar can be obtained by contracting the curvature tensor as
\begin{eqnarray}\nonumber
\tilde{\mathcal{C}}&=&\tilde{\mathcal{C}}^{\mu\nu} _{\mu\nu}
=\mathcal{C}+6\nabla_{\nu}h^{\nu}-4\nabla_{\nu}
\mathcal{T}^{\nu}-6h_{\nu}h^{\nu}
+8h_{\nu}\mathcal{T}^{\nu}+\mathcal{T}^{\mu\lambda\nu}
\mathcal{T}_{\mu\lambda\nu}
\\\label{14}
&+&2\mathcal{T}^{\mu\lambda\nu}\mathcal{T} _{\nu\lambda\mu}
-4\mathcal{T}^{\nu}\mathcal{T}_{\nu},
\end{eqnarray}
where$ \mathcal{T}_{\nu} = \mathcal{T}^{\nu}_{\mu\nu}$ and all
covariant derivatives are considered corresponding to the metric.

The gravitational action can be reformulated by eliminating the
boundary terms in the Ricci scalar as \cite{28}
\begin{equation}\label{15}
\mathcal{I}=\frac{1}{2\kappa} \int
\mathrm{g}^{\mu\nu}(\Gamma^{\lambda}_{\xi\mu}\Gamma^{\xi}_{\lambda\nu}
-\Gamma^{\lambda}_{\xi\lambda}\Gamma^{\xi}_{\mu\nu})\sqrt{-\mathrm{g}}
d^ {4}x.
\end{equation}
Based on the assumption that the connection is symmetric, we have
\begin{equation}\label{16}
\Gamma^{\lambda}_{\mu\nu}=-\mathrm{L}^{\lambda}_{\mu\nu}.
\end{equation}
Thus the gravitational action becomes
\begin{equation}\label{17}
\mathcal{I}=\frac{1}{2\kappa} \int
-\mathrm{g}^{\mu\nu}(\mathrm{L}^{\lambda}_{\xi\mu}
\mathrm{L}^{\xi}_{\lambda\nu} - \mathrm{L}^{\lambda}_{\xi\lambda}
\mathrm{L}^{\xi}_{\mu\nu}) \sqrt{-\mathrm{g}} d^ {4}x,
\end{equation}
where
\begin{equation}\label{18}
\mathrm{Q}\equiv-\mathrm{g}^{\mu\nu}(\mathrm{L}^{\lambda}_{\xi\mu}
\mathrm{L}^{\xi}_{\lambda\nu}
-\mathrm{L}^{\lambda}_{\xi\lambda}\mathrm{L}^{\xi}_{\mu\nu}),
\end{equation}
with
\begin{equation}\label{19}
\mathrm{L}^{\lambda}_{\xi\mu}\equiv-\frac{1}{2}
\mathrm{g}^{\lambda\varsigma}
(\nabla_{\mu}\mathrm{g}_{\xi\varsigma}+\nabla_{\xi}
\mathrm{g}_{\varsigma\lambda}
-\nabla_{\varsigma}\mathrm{g}_{\xi\mu}).
\end{equation}
From Eq.(\ref{17}), one can obtain the gravitational action of
$f(\mathrm{Q})$ theory by replacing non-mitricity scalar with an
arbitrary function as
\begin{equation}\label{20}
\mathcal{I}=\frac{1}{2\kappa}\int
\sqrt{-\mathrm{g}}f(\mathrm{Q})d^{4}x.
\end{equation}
This is the action of symmetric teleparallel theory, which is a
theoretical framework that provides an alternative geometric
description of gravity.

Now, we extend this gravitational Lagrangian by introducing the
trace of energy-momentum tensor in the functional action as
\begin{equation}\label{21}
\mathcal{I}=\frac{1}{2\kappa}\int f(\mathrm{Q},\mathrm{T})
\sqrt{-\mathrm{g}}d^{4}x.
\end{equation}
The modified Einstein-Hilbert action of $f(\mathrm{Q},\mathrm{T})$
gravity is defined as \cite{12}
\begin{equation}\label{22}
\mathcal{I}=\frac{1}{2\kappa}\int f(\mathrm{Q},\mathrm{T})
\sqrt{-\mathrm{g}}d^{4}x+\int
\mathrm{L}_{m}\sqrt{-\mathrm{g}}d^{4}x.
\end{equation}
The non-mitricity scalar is defined as
\begin{eqnarray}\label{23}
\mathrm{Q}_{\lambda}\equiv \mathrm{Q}^{~\mu}_{\lambda~~\mu}, \quad
\tilde{\mathrm{Q}}_{\lambda}\equiv \mathrm{Q}^{\mu}_{\lambda\mu}.
\end{eqnarray}
The superpotential of this model is given by
\begin{equation}\label{24}
\mathrm{P}^{\lambda}_{\mu\nu}=-\frac{1}{2}\mathrm{L}
^{\lambda}_{\mu\nu} +\frac{1}{4}(\mathrm{Q}^{\lambda}
-\tilde{\mathrm{Q}}^{\lambda})\mathrm{g}_{\mu\nu}- \frac{1}{4}
\delta ^{\lambda} _{(\mu \mathrm{Q}_{\nu})},
\end{equation}
and the relation for $\mathrm{Q}$ is
\begin{equation}\label{25}
\mathrm{Q}=-\mathrm{Q}_{\lambda\mu\nu}\mathrm{P}
^{\lambda\mu\nu}=-\frac{1}{4} (-\mathrm{Q}^{\lambda\nu\xi}
\mathrm{Q}_{\lambda\nu\xi}+2\mathrm{Q}^{\lambda\nu\xi}
\mathrm{Q}_{\xi\lambda\nu}
-2\mathrm{Q}^{\xi}\tilde{\mathrm{Q}}_{\xi}+\mathrm{Q}
^{\xi}\mathrm{Q}_{\xi}).
\end{equation}
The calculation of the above relation is shown in Appendix
\textbf{A}. By varying Eq.(\ref{22}) with respect to the metric
tensor, we obtain
\begin{eqnarray}\nonumber
\delta\mathcal{I}&=&\int \frac{1}{2\kappa} \delta
[f(\mathrm{Q},\mathrm{T}) \sqrt{-\mathrm{g}}] + \delta
[\mathrm{L}_{m} \sqrt{-\mathrm{g}}] d^ {4}x,
\\\nonumber
&=&\int \frac{1}{2\kappa}\big[-\frac{1}{2}f\mathrm{g}_{\mu\nu}
\sqrt{-\mathrm{g}} \delta \mathrm{g}^{\mu\nu} + f_{\mathrm{Q}}
\sqrt{-\mathrm{g}} \delta \mathrm{Q} + f_{\mathrm{T}}
\sqrt{-\mathrm{g}} \delta \mathrm{T}\big]
\\\label{26}
&-&\frac{1}{2} \mathrm{T}_{\mu\nu} \sqrt{-\mathrm{g}} \delta
\mathrm{g}^{\mu\nu}d^ {4}x.
\end{eqnarray}
The explicit formulation of $\delta\mathrm{Q}$ is given in Appendix
\textbf{B}. Moreover, we define
\begin{eqnarray}\label{27}
\mathrm{T}_{\mu\nu} \equiv \frac{-2}{\sqrt{-\mathrm{g}}}
\frac{\delta (\sqrt{-\mathrm{g}} \mathrm{L}_{m})}{\delta
\mathrm{g}^{\mu\nu}}, \quad \Theta_{\mu\nu} \equiv
\mathrm{g}^{\lambda\xi} \frac{\delta \mathrm{T}_{\lambda\xi}}{\delta
\mathrm{g}^{\mu\nu}},
\end{eqnarray}
which implies that $ \delta \mathrm{T}= \delta
(\mathrm{T}_{\mu\nu}\mathrm{g}^{\mu\nu}) = (\mathrm{T}_{\mu\nu}+
\Theta_{\mu\nu})\delta \mathrm{g}^{\mu\nu}$. Thus, Eq.(\ref{26})
turns out to be
\begin{eqnarray}\nonumber
\delta\mathcal{I}&=&\int \frac{1}{2\kappa}\bigg[\frac{-1}{2}f
\mathrm{g}_{\mu\nu}\sqrt{-\mathrm{g}} \delta \mathrm{g}^{\mu\nu} +
f_{\mathrm{T}}(\mathrm{T}_{\mu\nu}+
\Theta_{\mu\nu})\sqrt{-\mathrm{g}} \delta \mathrm{g}^{\mu\nu}
\\\nonumber
&-&f_{\mathrm{Q}} \sqrt{-\mathrm{g}} (\mathrm{P}_{\mu\lambda\xi}
\mathrm{Q}_{\nu}^{\lambda\xi}- 2\mathrm{Q}^{\lambda\xi} _{\mu}
\mathrm{P}_{\lambda\xi\nu}) \delta
\mathrm{g}^{\mu\nu}+2f_{\mathrm{Q}} \sqrt{-\mathrm{g}}
\mathrm{P}_{\lambda\mu\nu} \nabla^{\lambda} \delta
\mathrm{g}^{\mu\nu}
\\\label{28}
&-&\kappa\mathrm{T}_{\mu\nu}\sqrt{-\mathrm{g}} \delta
\mathrm{g}^{\mu\nu}\bigg]d^ {4}x.
\end{eqnarray}
After equating the variation of this equation to zero, the field
equations of the $f(\mathrm{Q},\mathrm{T})$ gravity turn out to be
\begin{eqnarray}\nonumber
\mathrm{T}_{\mu\nu}&=& \frac{-2}{\sqrt{-\mathrm{\mathrm{g}}}}
\nabla_{\lambda} (f_{\mathrm{Q}}\sqrt{-\mathrm{g}}
\mathrm{P}^{\lambda}_{\mu\nu})- \frac{1}{2} f \mathrm{g}_{\mu\nu} +
f_{\mathrm{T}} (\mathrm{T}_{\mu\nu} + \Theta_{\mu\nu})
\\\label{29}
&-&f_{\mathrm{Q}} (\mathrm{P}_{\mu\lambda\xi}
\mathrm{Q}_{\nu}^{\lambda\xi} -2\mathrm{Q}^{\lambda\xi}_{\mu}
\mathrm{P}_{\lambda\xi\nu}),
\end{eqnarray}
where $f_{\mathrm{T}}$ represents the derivative corresponding to
the trace of the energy-momentum tensor and $f_{\mathrm{Q}}$ defines
the derivative with respect to non-metricity. Equation (\ref{29})
represents the field equations in the context of the
$f(\mathrm{Q},\mathrm{T})$ theory, and the solution of these
equations can provide insights into how gravity behaves in this
modified framework.

\section{Field Equations and Karmarkar Condition}

We consider the inner region as
\begin{equation}\label{30}
ds^{2}=dt^{2}e^{\alpha(r)}-dr^{2}e^{\beta(r)}-d
\theta^{2}r^{2}-d\phi^{2}r^{2}\sin^{2}\theta.
\end{equation}
The stress-energy tensor manifests configurations of matter and
energy in a system and its components yield physical features that
produce distinct aspects of its dynamics. We consider anisotropic
matter distribution as
\begin{equation}\label{31}
\mathrm{T}_{\mu\nu}=\mathrm{U}_{\mu}\mathrm{U}_{\nu}\varrho +
\mathrm{V}_{\mu}\mathrm{V}_{\nu}p_{r}-p_{t}\mathrm{g}_{\mu\nu}+
\mathrm{U}_{\mu}\mathrm{U}_{\nu}p_{t}
-\mathrm{V}_{\mu}\mathrm{V}_{\nu}p_{t},
\end{equation}
where four-vector and four-velocity of the fluid are denoted by
$\mathrm{V}_{\mu}$ and $\mathrm{U}_{\mu}$, respectively. In the
context of gravitational physics, the matter-Lagrangian density is a
fundamental concept that describes the distribution of matter and
its dynamics in a given spacetime. We consider matter-Lagrangian
density for anisotropic matter as
$\mathrm{L}{m}=-\frac{p{r}+2p_{t}}{3}$ \cite{29}. This Lagrangian
density has been employed in various theoretical models and has
shown reasonable success in explaining observations and phenomena
involving anisotropic matter distributions, especially in the
context of CSs and certain cosmological scenarios. The corresponding
field equations turn out to be
\begin{eqnarray}\nonumber
\varrho&=&\frac{1}{2r^{2}e^{\beta}}\bigg[2r\mathrm{Q}'f_{\mathrm{Q}
\mathrm{Q}}(e^{\beta}-1)
+f_{\mathrm{Q}}\big((e^{\beta}-1)(2+r\alpha')+(e^{\beta}+1)r\beta'
\big)
\\\label{33}
&+&fr^{2}e^{\beta}\bigg]-\frac{1}{3}f_{\mathrm{T}}(3\varrho+p_{r}+2p_{t}),
\\\nonumber
p_{r}&=&\frac{-1}{2r^{2}e^{\beta}}\bigg[2r\mathrm{Q}'f_{\mathrm{Q}
\mathrm{Q}}(e^{\beta}-1)
+f_{\mathrm{Q}}\big((e^{\beta}-1)(2+r\alpha'+r\beta')-2r\alpha'\big)
\\\label{34}
&+&fr^{2}e^{\beta}\bigg]+\frac{2}{3}f_{\mathrm{T}}(p_{t}-p_{r}),
\\\nonumber
p_{t}&=&\frac{-1}{4re^{\beta}}\bigg[-2r\mathrm{Q}'\alpha'f_{\mathrm{Q}
\mathrm{Q}}
+f_{\mathrm{Q}}\big(2\alpha'(e^{\beta}-2)-r\alpha'^{2}
+\beta'(2e^{\beta}+r\alpha')
\\\label{35}
&-&2r\alpha''\big)+2fre^{\beta}\bigg]+\frac{1}{3}f_{\mathrm{T}}
(p_{r}-p_{t}).
\end{eqnarray}

When dealing with multivariate functions and their derivatives, the
resulting field equations become more intricate, making it
challenging to derive specific outcomes. Consequently, to facilitate
analysis, we opt for a particular model of
$f(\mathrm{Q},\mathrm{T})$ as \cite{30}
\begin{eqnarray}\label{36}
f(\mathrm{Q},\mathrm{T})=\gamma\mathrm{Q}+\eta\mathrm{T},
\end{eqnarray}
where $\gamma$ and $\eta$ are arbitrary constants.

The key features of this model lie in its departure from the
standard Einstein-Hilbert action, allowing for a more comprehensive
description of gravitational interactions. By incorporating the
trace of the stress-energy tensor, the theory captures non-trivial
effects arising from the energy-momentum distribution, which is
particularly relevant in the context of CSs where extreme densities
and pressures prevail. The assumptions underlying this model include
the validity of the Einstein equivalence principle, the existence of
a metric theory and the requirement of covariance under general
coordinate transformations. These assumptions enable the formulation
of field equations governing the dynamics of matter and geometry in
the framework of $f(\mathrm{Q},\mathrm{T})$ gravity. The
implications of employing the $f(\mathrm{Q},\mathrm{T})$ theory in
our study manifest in the altered field equations and subsequently
in the behavior of matter and geometry in CSs. Specifically,
deviations from the predictions of EGTR arise in regions of high
energy density, impacting the structure and properties of CSs.
Therefore, by elucidating the specific model and its implications,
we aim to provide a clearer understanding of the interplay between
gravitational interactions and the properties of CSs in the
framework of $f(\mathrm{Q},\mathrm{T})$ gravity. The considered
cosmological model has been widely used in the literature \cite{31}.

This model of the $f(\mathrm{Q},\mathrm{T})$ theory emerges as a
crucial endeavor in the realm of theoretical physics, particularly
in the pursuit of understanding the fundamental nature of physical
phenomena at both macroscopic and microscopic scales. Motivated by
the quest for a unified framework that can elegantly encapsulate
diverse phenomena spanning from cosmology to particle physics, the
corresponding model of $f(\mathrm{Q},\mathrm{T})$ theory aims to
provide a comprehensive and insightful description of the universe's
behavior. One of the primary motivations behind the development of
this model lies in addressing the limitations and gaps present in
existing theoretical frameworks such as EGTR and quantum mechanics.
While these theories have been remarkably successful in explaining
various aspects of the universe, they encounter challenges when
confronted with scenarios involving extreme conditions or unexplored
domains, such as the early universe or black hole singularities.

This model allows for the incorporation of additional degrees of
freedom and geometric structures, thereby enabling a more nuanced
description of gravitational interactions and their interplay with
matter and energy. Furthermore, the minimal model of
$f(\mathrm{Q},\mathrm{T})$ theory is motivated by its potential to
reconcile discrepancies observed between theoretical predictions and
experimental observations. By refining the mathematical formalism
and introducing novel dynamical mechanisms, this framework aims to
provide more accurate predictions for phenomena such as dark energy,
dark matter and gravitational waves, which remain enigmatic in the
current theoretical paradigm. Moreover, the quest for a minimal
model of $f(\mathrm{Q},\mathrm{T})$ theory is intrinsically linked
to broader theoretical endeavors, including the quest for a theory
of quantum gravity and the exploration of the fundamental nature of
spacetime itself. By probing the intricate interplay between
geometry and matter in the context of $f(\mathrm{Q},\mathrm{T})$
theory, researchers hope to uncover deeper insights into the fabric
of the universe and its underlying symmetries. The motivation behind
this model of $f(\mathrm{Q},\mathrm{T})$ theory stems from its
potential to overcome existing theoretical limitations, reconcile
theoretical predictions with experimental observations and provide a
more comprehensive understanding of the fundamental laws governing
the universe. By embracing this framework, physicists embark on a
journey towards unraveling the mysteries of the cosmos and unlocking
the secrets of nature at its most fundamental level.

The resulting modified field equations are
\begin{eqnarray}\nonumber
\varrho&=&\frac{\gamma e^{-\beta}}{12r^2(2\eta^{2}+\eta-1)}\bigg[
\eta(2r(-\beta'(r\alpha'+2)+2r\alpha''+\alpha'(r\alpha'+4))-4e^{\beta}
\\\label{37}
&+&4)+3\eta r(\alpha'(4-r\beta'+r\alpha')+2r\alpha'')+12
(\eta-1)(r\beta'+e^{\beta}-1)\bigg],
\\\nonumber
p_{r}&=&\frac{\gamma e^{-\beta}}{12r^2(2\eta^{2}+\eta-1)}\bigg[
2\eta\big(r\beta'(r\alpha'+2)+2(e^{\beta}-1)-r(2r\alpha''+\alpha'(r\alpha'
\\\nonumber
&+&4))\big)+3\big(r\big(\eta\beta'(r\alpha'+4)-2\eta
r\alpha''-\alpha'(-4\eta+\eta r\alpha '+4)\big)-4(\eta-1)
\\\label{38}
&\times&(e^{\beta}-1)\big)\bigg],
\\\nonumber
p_{t}&=&\frac{\gamma e^{-\beta}}{12r^2(2\eta^{2}+\eta-1)}\bigg[
2\eta\big(r\beta'(r\alpha'+2)+2(e^{\beta}-1)-r(2r\alpha''+\alpha'
\\\nonumber
&\times&(r\alpha'+4))\big)+3\big(r
\big(2(\eta-1)r\alpha''-((\eta-1)r\alpha'-2)(\beta'-\alpha')\big)
\\\label{39}
&+&4\eta(e^{\beta}-1)\big)\bigg].
\end{eqnarray}
The well-known Karmarkar condition is defined as \cite{32}
\begin{eqnarray}\label{40}
\mathrm{R}_{1414}\mathrm{R}_{2323}=\mathrm{R}_{1212}\mathrm{R}_{3434}
+\mathrm{R}_{1224}\mathrm{R}_{1334},
\end{eqnarray}
where
\begin{eqnarray}\nonumber
\mathrm{R}_{1414}&=&-\frac{1}{4}(2\alpha''+\alpha'^{2}-\alpha'\beta')e^{\alpha},
\quad
\mathrm{R}_{2323}=\frac{1}{e^{\beta}}r^{2}\sin^{2}\theta(e^{\beta}-1),
\\\nonumber
\mathrm{R}_{3434}&=&\frac{-1}{2}r^{2}\sin^{2}\theta
\alpha'e^{\alpha-\beta}, \quad \mathrm{R}_{1212}=\frac{1}{2}r\beta',
\quad \mathrm{R}_{1334}=\mathrm{R}_{1224}\sin^{2}\theta=0.
\end{eqnarray}
This condition was formulated by the Indian mathematician and
physicist Shanti Swarup Karmarkar in the 1940s. Its applications
extend to diverse fields, including the derivation of precise
solutions, exploration of singularity theorems and the study of
cosmological models. It is crucial to recognize that this condition
serves as a mathematical instrument for comprehending specific
properties of spacetime. It is a geometric condition that involves
the Riemann curvature tensors, which characterize the curvature of
spacetime. This condition plays a crucial role in understanding the
behavior of spacetime in the presence of matter and energy. This
constraint has been extensively explored in various contexts such as
exact solutions and cosmological models. It is essential to
recognize that the Karmarkar condition serves as a specific
mathematical tool, facilitating the comprehension of certain
spacetime properties. This condition is expressed mathematically as
an equation that relates the metric coefficients in the
four-dimensional spacetime to certain parameters in the
higher-dimensional space. The application of the Karmarkar condition
is a technique to generate solutions for the metric functions
describing the geometry of a CS. Once these solutions are obtained,
further analysis is performed to assess the physical properties of
the CSs such as their stability and the matter content required to
sustain it. It is important to note that the Karmarkar condition is
an approach among various methods used to study CSs and
gravitational solutions. The motivation for considering the
Karmarkar condition in stellar structure lies in its ability to
provide insights into the stability of stars. By analyzing this
condition, astronomers can assess whether a star's structure will
remain stable over time or if it's prone to instabilities that could
lead to phenomena like collapse or disruption.

The Karmarkar condition provides a valuable framework for
understanding the equilibrium conditions in stellar interiors,
particularly in the context of gravitational collapse and stability
analysis. By incorporating the Karmarkar condition into our
analysis, we aim to enhance our understanding of the physical
processes governing stellar evolution and structure. Specifically,
the advantage of considering this condition lies in its ability to
provide insights into the internal pressure-density distribution and
the overall stability of stellar configurations. Furthermore, the
Karmarkar condition offers a rigorous mathematical framework for
exploring the behavior of compact astrophysical objects such as
white dwarfs, neutron stars and black holes, under extreme
conditions of gravitational pressure. By elucidating the
implications of this condition on stellar models, we can better
constrain theoretical predictions and refine our interpretations of
observational data. Thus, the Karmarkar condition provides a
valuable tool for studying the stability of stellar structures,
helping astronomers better understand the dynamics and evolution of
stars. A lot of work based on the Karmarkar condition in the
framework of different modified theories has been done in
\cite{32b}-\cite{32j}.

Solving Eq.(\ref{40}), we obtain
\begin{equation}\label{41}
\alpha'(\beta'-\alpha')-2\alpha''=\frac{\alpha'\beta'}{1-e^{\beta}},
\end{equation}
where $e^{\beta}\neq 1$. Integrating this equation, we have
\begin{equation}\label{42}
e^{\alpha(r)}=(c+d \int\sqrt{e^{\beta}-1}dr)^{2}.
\end{equation}
Here, the integration constants are denoted by $c$ and $d$. Further,
we choose the metric potential $e^{\beta}$ as \cite{34}
\begin{equation}\label{43}
e^{\beta(r)}=1+(1 + b r^{2})^{\lambda}a^{2} r^{2},
\end{equation}
where one can select $\lambda$ as any real number except zero. In
the framework of EGTR, Singh and Pant \cite{33} considered positive
values of $\lambda$ to analyze the physical characteristics of
compact stellar objects. Later, Bhar et al \cite{34} examined this
metric element with $\lambda=-4$ and discovered stable cosmic
objects. Inspired by their work, we aim to extend the analysis in
the context of $f(Q,T)$ gravity by choosing $\lambda=-4$. However,
we attempted other choices of $\lambda$ (both positive and negative)
but could not obtain viable results in these cases.

By manipulating Eqs.(\ref{42}) and (\ref{43}), we obtain
\begin{eqnarray}\label{44}
e^{\alpha(r)}=\bigg(c -\frac{ad}{2b(1+br^{2})}\bigg)^{2}.
\end{eqnarray}
The unknown constants $(a, b, c, d)$ can be found using the first
Darmois junction condition. This condition matches interior and
exterior spacetimes at the hypersurface. By imposing this condition,
researchers can model the behavior of matter in celestial objects,
leading to a deeper understanding of their physical properties. We
consider the outer geometry of the CSs as
\begin{eqnarray}\label{45}
ds^{2}_{+}=dt^{2}(1-\frac{2m}{r})-dr^{2}(1-\frac{2m}{r})^{-1}-d\theta^{2}r^{2}
-d\phi^{2}r^{2}\sin^{2}\theta.
\end{eqnarray}
The continuity of metric coefficients of the metrics (\ref{30}) and
(\ref{45}) at the surface boundary $(r=\mathcal{R})$ gives
\begin{eqnarray}\label{46}
\mathrm{g}_{tt}&=&1+\frac{a^{2}r^{2}}{(1+br^{2})^{4}}=
1-\frac{2m}{\mathcal{R}},
\\\label{47}
\mathrm{g}_{rr}&=&\big(c-\frac{ad}{2b(1+br^{2})}\big)^{2}=
(1-\frac{2m}{\mathcal{R}})^{-1},
\\\label{48}
\mathrm{g}_{tt,r}&=&\frac{2adr\big(c-\frac{ad}{2b(1+br^{2})}
\big)}{(1+br^2)^2}=\frac{2m}{\mathcal{R}^{2}}.
\end{eqnarray}
By solving the above equations simultaneously, we obtain
\begin{eqnarray}\label{49}
a&=&\frac{(1+b\mathcal{R}^2)^2}{\mathcal{R}}
\big(\frac{2m}{\mathcal{R}-2m}\big)^{1/2},
\\\label{50}
b&=&\frac{(4m-\mathcal{R})}{\mathcal{R}^2(9\mathcal{R}-20m)},
\\\label{51}
c&=&\frac{ad}{2b(1+b\mathcal{R}^2)}+\big(1-\frac{2m}{\mathcal{R}}\big)^{1/2},
\\\label{52}
d&=&\frac{1}{2\mathcal{R}}\big(\frac{2m}{\mathcal{R}}\big)^{1/2}.
\end{eqnarray}
\begin{table}\caption{Approximate values of input parameters.}
\begin{center}
\begin{tabular}{|c|c|c|}
\hline Compact stars & $M_{\odot}$ & $\mathcal{R}(km)$
\\
\hline  4U 1538-52 \cite{35} & 0.87 $\pm$ 0.07 & 7.866 $\pm$ 0.21
\\
\hline  SAX J1808.4-3658 \cite{36} & 0.9 $\pm$ 0.3 & 7.951 $\pm$ 1.0
\\
\hline  Her X-1 \cite{37} & 0.85 $\pm$ 0.15 & 8.1 $\pm$ 0.41
\\
\hline  LMC X-4  \cite{35} & 1.04 $\pm$ 0.09 & 8.301 $\pm$ 0.2
\\
\hline  EXO 1785-248 \cite{38} & 1.30 $\pm$ 0.2 & 10.10 $\pm$ 0.44
\\
\hline
\end{tabular}
\end{center}
\end{table}
\begin{table}\caption{Approximate values of output parameters.}
\begin{center}
\begin{tabular}{|c|c|c|c|c|}
\hline Compact stars & $a$ & $b$ & $c$ & $d$
\\
\hline  4U 1538-52 & 0.0780314 & -0.000979628 & -0.717909 &
0.0362963
\\
\hline  SAX J1808.4-3658 & 0.0788571 & -0.000929059 & -0.821585 &
0.0363264
\\
\hline  Her X-1 & 0.072303 & -0.000983907 & -0.517508 & 0.0343333
\\
\hline  LMC X-4 & 0.0833349 & -0.000714639 & -1.45075 & 0.0366062
\\
\hline  EXO 1785-248  & 0.0704159 & -0.000454051 & -1.6917 &
0.0304946
\\
\hline
\end{tabular}
\end{center}
\end{table}
\begin{figure}
\epsfig{file=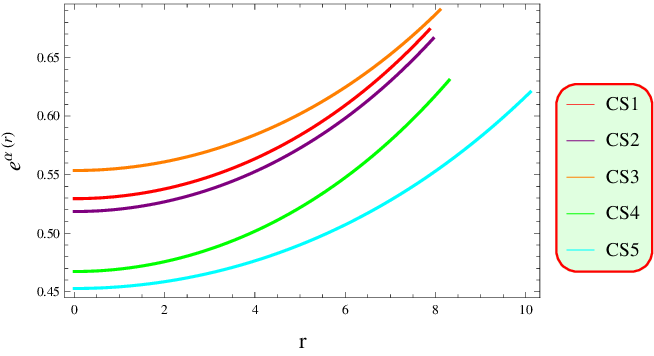,width=.5\linewidth}
\epsfig{file=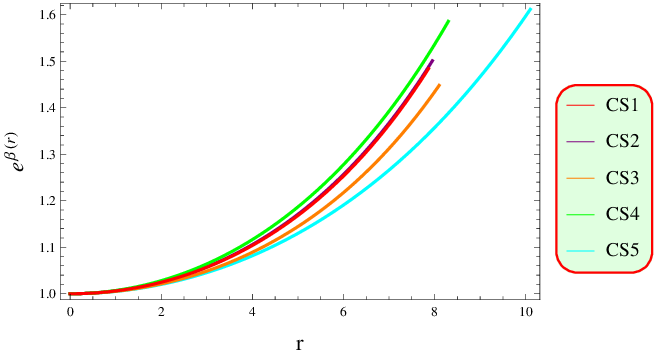,width=.5\linewidth}\caption{Graphs of metric
potential versus radial coordinate.}
\end{figure}

The Karmarkar solutions describe the gravitational field around CSs.
These solutions involve integration constants that provide essential
insights into the gravitational and physical properties of CSs,
aiding in the understanding of their structure. These solutions
describe the distribution of energy in the CSs. For example, in the
case of a neutron star, it provides insight into how the energy is
distributed in the interior, which is crucial for understanding its
stability and structural properties. The solutions can be used to
understand gravitational redshift effects near the surface of CSs.
Due to the intense gravitational field, electromagnetic radiation
emitted from the surface experiences a significant redshift as it
travels away from the objects. Depending on the integration
constants of the Karmarkar solutions, it may contain features such
as event horizons and singularities. These features are indicative
of the formation of black holes under certain conditions, providing
insight into the final stages of stellar collapse. Thus, the
integration constants in Karmarkar solutions offer a theoretical
framework for studying the gravitational effects and structural
properties of compact stellar objects. Their physical interpretation
provides valuable insights into the nature of these objects and
their observable characteristics.

Table \textbf{1} presents the observed values of mass and radius for
the considered CSs, while Table \textbf{2} contains the
corresponding constants associated with mass and radius. The
solution's compatibility is guaranteed by the non-singular and
positively increasing behavior of metric elements across the entire
domain. Figure \textbf{1} determine the behavior of metric
potentials, confirming the regularity and necessary positive
increase of both metric elements. In the graphs, the abbreviations
CS1, CS2, CS3, CS4, and CS5 refer to 4U 1538-52, SAX J1808.4-3658,
Her X-1, LMC X-4, and EXO 1785-248 CSs, respectively. The
corresponding field equations are
\begin{eqnarray}\nonumber
\varrho&=&-a\gamma\bigg[10bd(br^2-3)(1+br^2)^{6}\eta+a^{4}
dr^2(2\eta-3)-2a^{3}cbr^{2}(1+br^2)
\\\nonumber
&\times&(2\eta-3)+2acb(1+br^{2})^{4}(5br^{2}-3)(2\eta-3)
-3a^{2}d(br^{2}+1)^{3}(3-2\eta
\\\nonumber
&+&5br^{2}(2\eta-1))\bigg]\bigg[3(-ad+2cb(1+br^{2}))(a^{2}
r^{2}+(1+br^{2})^{4})^{2}(1+\eta)
\\\label{53}
&\times&(2\eta-1)\bigg]^{-1},
\\\nonumber
p_{r}&=&-a\gamma\bigg[a^{4}r^{2}d(3-2\eta)+2a^{3}cbr^{2}
(1+br^{2})(2\eta-3)-3a^{2}d(1+br^{2})^{3}
\\\nonumber
&\times&(-1-2\eta+5br^{2}(2\eta-1))-2bd(1+br^{2})^{6}
(-3(2+\eta)+br^{2}(17\eta-6))
\\\nonumber
&+&2acb(1+br^{2})^{4}(-3-6\eta+br^{2}(26\eta-3))\bigg]
\bigg[3(-ad+2cb(1+br^{2}))(a^{2}r^{2}
\\\label{54}
&+&(1+br^{2})^{4})^{2}(1+\eta)(2\eta-1)\bigg]^{-1},
\\\nonumber
p_{t}&=&-a\gamma\bigg[4a^{4}dr^{2}\eta-8a^{3}cbr^{2}
(1+br^{2})\eta+3a^{2}d(1+br^{2})^{3}(1+2\eta
+br^{2}
\\\nonumber
&\times&(2\eta-1))+2acb(1+br^{2})^{4}(-3-6\eta+br^{2}
(9+2\eta))+2b
d(1+br^{2})^{6}
\\\nonumber
&\times&(3(2+\eta)+br^{2}(7\eta-6))\bigg]\bigg[3(2cb-ad
(1+br^{2}))(a^{2}r^{2}+(1+br^{2})^{4})^{2}
\\\label{55}
&\times&(1+\eta)(2\eta-1)\bigg]^{-1}.
\end{eqnarray}

\section{Viable Features of Compact Stars}

In this section, we analyze the viable characteristics of CSs and
examine their behavior graphically. Our investigation delves into
the influence of various physical parameters such as material
variables, energy constraints, EoS parameters, mass, compactness,
redshift, equilibrium state (as governed by the Tolman-Oppenheimer
equation) and stability analysis (in terms of sound speed and
adiabatic index) through graphical representations.

\subsection{Energy Density and Pressure Components}

The study of fluid variables such as energy density, radial pressure
and tangential pressure in self-gravitating stellar objects is
crucial for understanding their internal structure. These variables
are expected to maximum at the core due to the dense profile of
these objects. The positive behavior counteracts the gravitational
force, providing support against collapse. Figures
\textbf{2}-\textbf{3} represent the graphical representation of
fluid parameters and their derivatives for each star candidate.
These characteristics reach their maximum at the center and exhibit
a positive decrease, indicating a highly compact profile for the
proposed CSs. Additionally, the radial pressure in each candidate
consistently decreases as the radial coordinate increases,
ultimately vanishing at the boundary. Figure \textbf{3} further
demonstrates that the derivative of fluid parameters is zero at the
center, turning negative, thus confirming the presence of a highly
compact configuration in the context of $f(\mathrm{Q},\mathrm{T})$
theory.
\begin{figure}
\epsfig{file=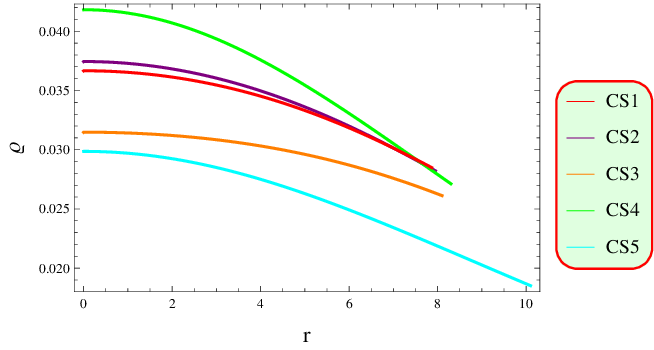,width=.5\linewidth}
\epsfig{file=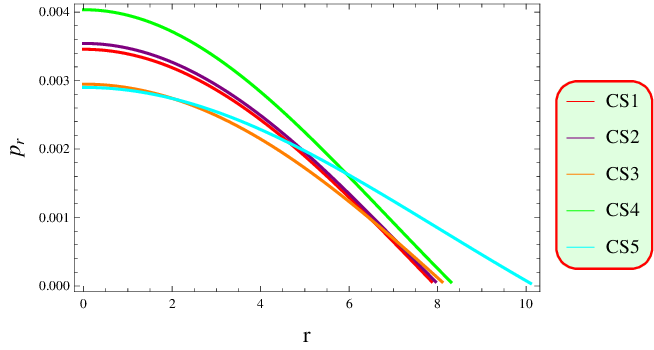,width=.5\linewidth}\center
\epsfig{file=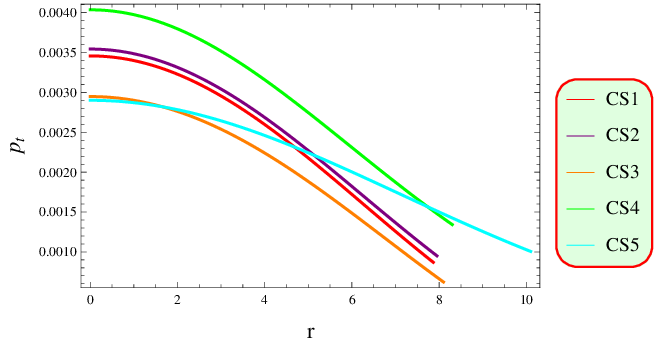,width=.5\linewidth}\caption{Evolution of matter
contents versus radial coordinate.}
\end{figure}
\begin{figure}
\epsfig{file=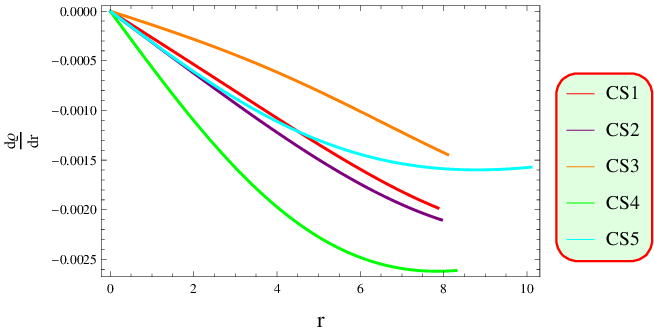,width=.5\linewidth}
\epsfig{file=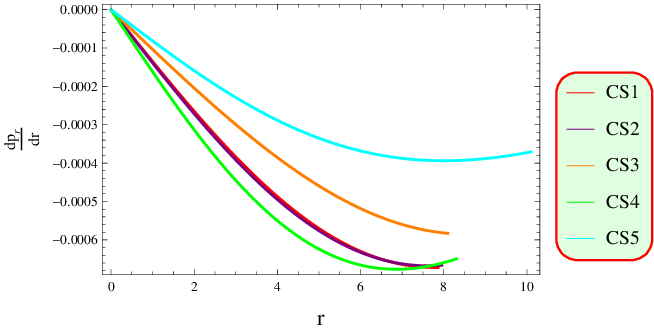,width=.5\linewidth}\center
\epsfig{file=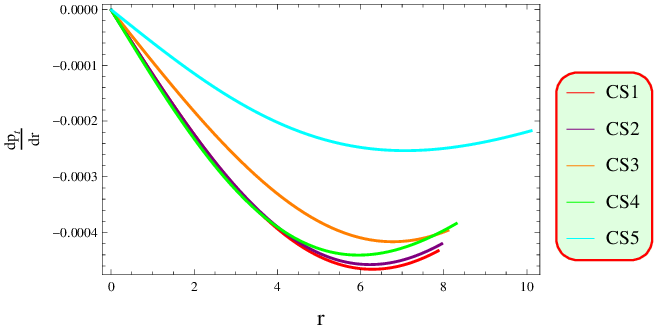,width=.5\linewidth}\caption{Evolution of
gradient of matter contents versus radial coordinate.}
\end{figure}

Pressure anisotropy $(\Delta = p_t - p_r)$ refers to the phenomenon
where pressure in a system is not uniform in all directions. In
other words, the pressure can vary depending on the direction in
which it is measured. Positive anisotropy implies outward pressure,
while negative anisotropy signifies inward pressure. Figure
\textbf{4} shows the anisotropic behavior of the proposed CSs. The
depicted trend reveals a consistent increase in anisotropy for all
CSs, indicating the presence of a repulsive force essential for
substantial geometries \cite{40}.
\begin{figure}\center
\epsfig{file=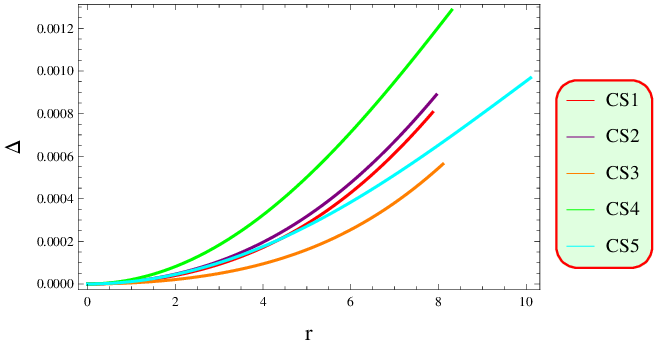,width=.5\linewidth}\caption{Behavior of
$p_t-p_r$ versus radial coordinate.}
\end{figure}

\subsection{Analysis of Normal Matter}

To explore the presence of feasible cosmic structures, it is
essential to impose particular constraints on matter, referred to as
energy conditions. These conditions encompass a series of
inequalities that set limitations on the stress-energy tensor,
governing the behavior of matter and energy under the influence of
gravity. Various energy conditions exist, each imposing distinct
restrictions on the fluid parameters as
\begin{itemize}
\item Null energy constraint
\begin{eqnarray}\nonumber
0\leq p_{r}+\varrho, \quad 0\leq p_{t}+\varrho.
\end{eqnarray}
\item Dominant energy constraint
\begin{eqnarray}\nonumber
0\leq \varrho\pm p_{r}, \quad 0\leq \varrho\pm p_{t}.
\end{eqnarray}
\item Weak energy constraint
\begin{eqnarray}\nonumber
0\leq p_{r}+\varrho,\quad 0\leq p_{t}+\varrho, \quad 0\leq \varrho.
\end{eqnarray}
\item Strong energy constraint
\begin{eqnarray}\nonumber
0\leq p_{r}+\varrho, \quad 0\leq p_{t}+\varrho, \quad 0\leq
p_{r}+2p_{t}+\varrho.
\end{eqnarray}
\end{itemize}
Viable cosmic structures must satisfy these conditions. By studying
the energy conditions, one can gain insights into the nature of
matter and energy, the viability of cosmic structures and the
possibilities of phenomena beyond our current understanding of
physics. Figure \textbf{5} demonstrates that matter inside the
cosmic structures is ordinary as all the energy constraints are
satisfied in the presence of $f(\mathrm{Q},\mathrm{T})$ terms.
\begin{figure}
\epsfig{file=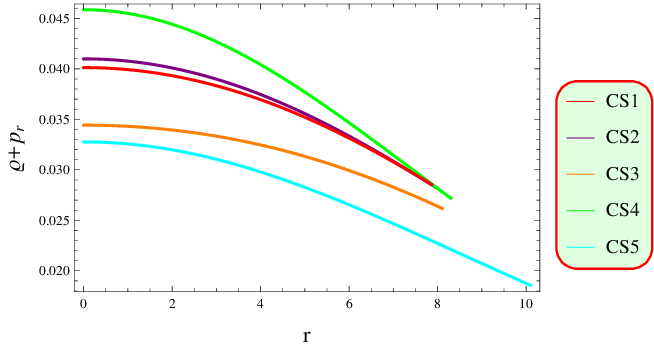,width=.5\linewidth}
\epsfig{file=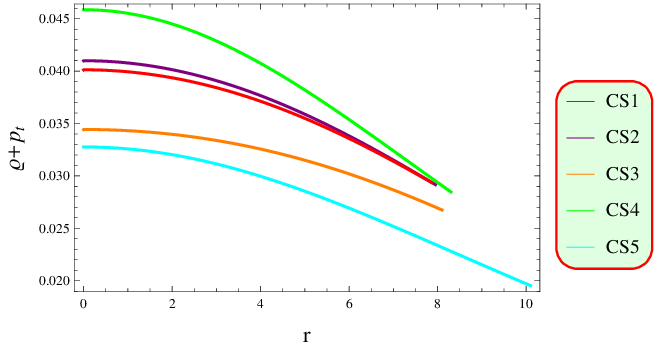,width=.5\linewidth}
\epsfig{file=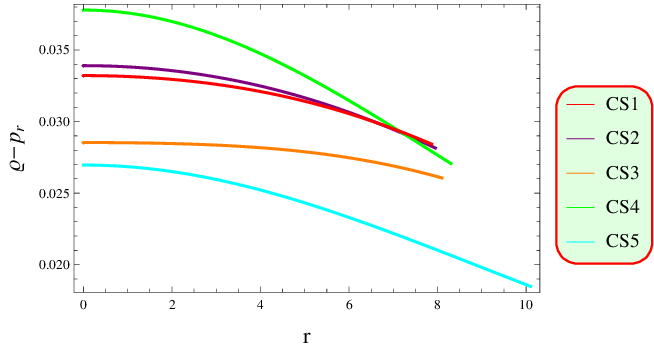,width=.5\linewidth}
\epsfig{file=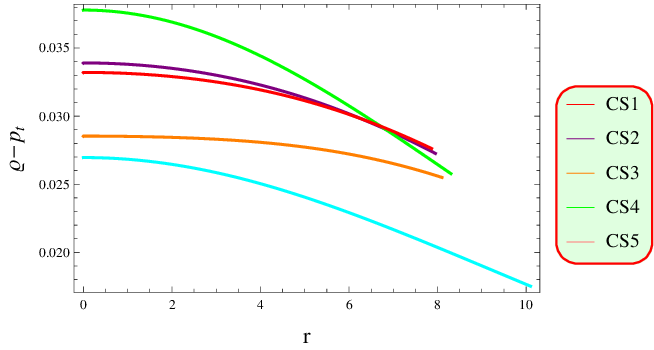,width=.5\linewidth}\caption{Behavior of energy
conditions versus radial coordinate.}
\end{figure}

\subsection{Analysis of State Parameters}

Here, we investigate the equation of state parameters that are
crucial in describing the relation between pressure and energy
density in various physical systems. The radial
$(\omega_{r}=\frac{p_{r}}{\varrho})$ and transverse
$(\omega_{t}=\frac{p_{t}}{\varrho})$ components must lie in [0,1]
for a physically viable model \cite{42}. Using
Eqs.(\ref{53})-(\ref{55}), we have
\begin{eqnarray}\nonumber
\omega_{r}&=&\bigg[a^{4}dr^{2}(3-2\eta)+2a^{3}cbr^{2}
(1+br^{2})(2\eta-3)-3a^{2}d(1+br^{2})^{3}
(-1-2\eta
\\\nonumber
&+&5br^{2}(2\eta-1))-2bc(1+br^{2})^{6}(-3(2+\eta)+br^{2}
(17\eta-6))+2acb
\\\nonumber
&\times&(1+br^{2})^{4}(-3-6\eta+br^{2}(26\eta-3))\bigg]
\bigg[10bd(br^{2}-3)(1+br^{2})^{6}\eta
\\\nonumber
&+&a^{4}dr^2(2\eta-3)-2a^{3}cbr^{2}(1+br^{2})(2\eta-3)
+2acb(1+br^{2})^{4}(5br^{2}-3)
\\\nonumber
&\times&(2\eta-3)-3a^{2}d(1+br^{2})^{3}(3-2\eta+5br^{2}
(2\eta-1))\bigg]^{-1},
\\\nonumber
\omega_{t}&=&\bigg[4a^{4}dr^{2}\eta-8a^{3}cbr^{2}(1+br^{2})
\eta+3a^{2}d(1+br^{2})^{3}(1+2\eta+br^{2}(2\eta-1))
\\\nonumber
&+&2acb(1+br^{2})^{4}(-3-6\eta+br^{2}(9+2\eta))+2bd
(1+br^{2})^{6}(3(2+\eta)
\\\nonumber
&+&br^{2}(7\eta-6))\bigg]\bigg[10bd(br^{2}-3)(1+br^{2})^{6}
\eta+a^{4}dr^{2}(2\eta-3)-2a^{3}cbr^{2}
\\\nonumber
&\times&(1+br^{2})(2\eta-3)+2acb(1+br^{2})^{4}(5br^{2}-3)
(2\eta-3)-3a^{2}d(1+br^{2})^{3}
\\\nonumber
&\times&(3-2\eta+5br^{2}(2\eta-1))\bigg]^{-1}.
\end{eqnarray}
The graphical analysis of EoS parameters is given in Figure
\textbf{6}, which determines that EoS parameters satisfy the
required viability condition of the considered CSs.
\begin{figure}
\epsfig{file=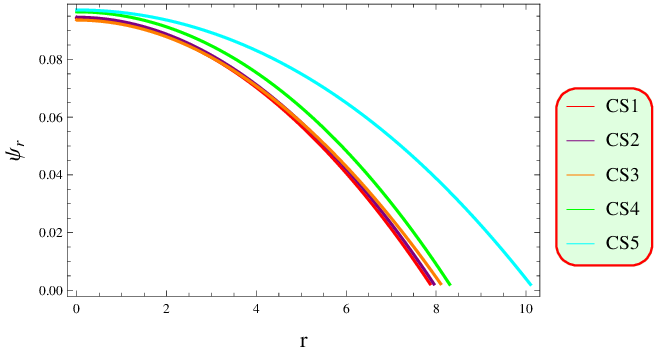,width=.5\linewidth}
\epsfig{file=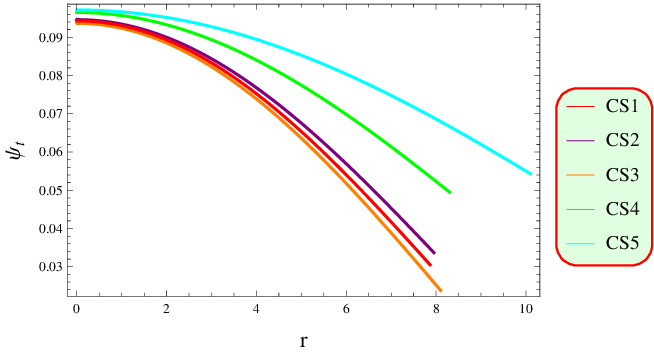,width=.5\linewidth}\caption{Graphs of radial
and tangential components of EoS parameter versus radial
coordinate.}
\end{figure}

\subsection{Evolution of Different Physical Aspects}

The mass of anisotropic CS is defined as
\begin{equation}\label{56}
M=4\pi\int^{\mathcal{R}}_{0} r^{2}\varrho dr.
\end{equation}
This equation is numerically solved with the initial condition
$M(0)=0$. Figure \textbf{7} manifests the graphical trend of mass
distribution in each CS based on this numerical solution. The graph
depicts a consistent and positive increase in mass as the radius
expands. Moreover, as the radius approaches to 0, the mass function
converges to 0, indicating a regular behavior at the center of CSs.
Analyzing the structural composition of celestial objects involves
exploring various physical aspects. One such aspect is the
compactness function, denoted as $u=\frac{M}{r}$ which plays a
pivotal role in assessing the viability of CSs. Buchdahl \cite{43}
introduced a significant limit for the mass-radius ratio, suggesting
that viable CSs should satisfy the condition $u<4/9$.

The surface redshift serves as a crucial parameter for studying the
characteristics of CSs, measuring the change in the wavelength of
emitted light due to strong gravitational influences. The expression
for surface redshift in terms of compactness is given by
\begin{equation}\label{57}
Z_s = -1 + \frac{1}{\sqrt{1-2u}}.
\end{equation}
Buchdahl \cite{43} established a critical condition stating that the
surface redshift must be less than 2 for viable CSs with a perfect
matter distribution. However, Ivanov \cite{44} reported a value of
5.211 for anisotropic configurations. Both the compactness and
redshift functions exhibit a monotonically increasing behavior,
reaching zero at the star's center as shown in Figure \textbf{8}. It
is noteworthy that both functions adhere to specified limits
($u<4/9$ and $Z_{s}<5.211)$.
\begin{figure}\center
\epsfig{file=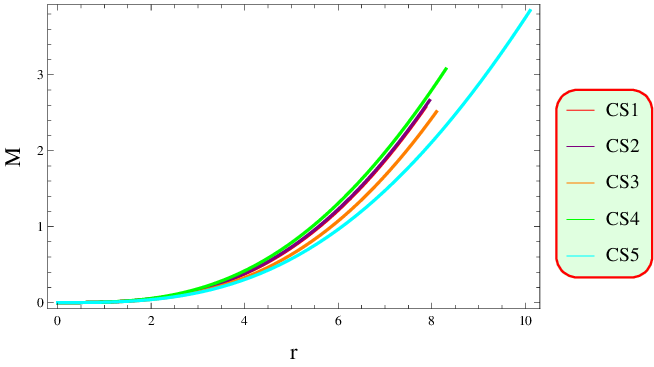,width=.5\linewidth}\caption{Plots of mass
versus radial coordinate.}
\end{figure}
\begin{figure}
\epsfig{file=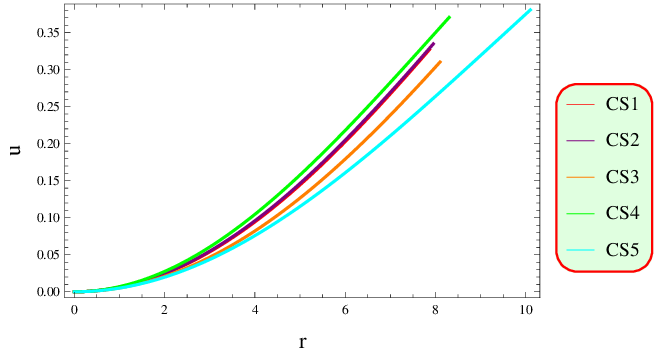,width=.5\linewidth}
\epsfig{file=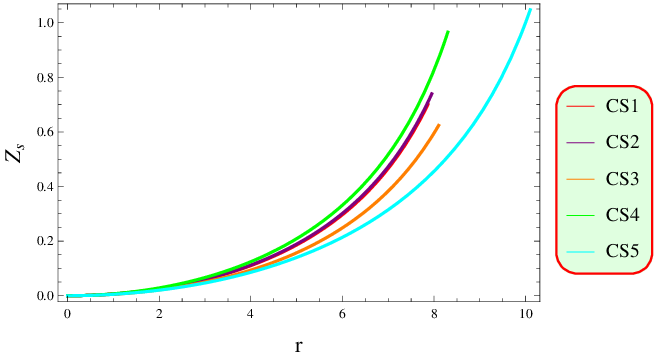,width=.5\linewidth}\caption{Graphs of
compactness factor and surface redshift versus radial coordinate.}
\end{figure}

\subsection{Mass-Radius and Mass-Inertia Relations}

Mass-Radius $(M-R)$ and Mass-Inertia $(M-I)$ relations are essential
tools in the study of CSs. These provide insights into the
relationship between mass, radius and moment of inertia for the
cosmic objects. The Mass-Radius relationship helps in understanding
how the mass of a CS influences its size and structure. The $M-R$
relation helps astronomers to compare theoretical models with
observational data, allowing them to constrain the properties of CSs
and understand their internal structure. The $M-I$ shows the
relationship between the mass and the moment of inertia of a CS.
Moment of inertia is a measure of an object's resistance to changes
in rotation. It depends not only on the mass distribution but also
on the object's shape and size. In the context of CSs, the $M-I$
relation provides insights into how mass affects the rotational
properties of these objects. For example, the neutron stars are
known for their rapid rotation. As the mass of a neutron star
increases, its moment of inertia also increases, affecting its
rotational behavior. The $M-I$ relation is very useful for studying
pulsars, which are rapidly rotating neutron stars emitting beams of
electromagnetic radiation. By observing the pulsar's rotation period
and understanding its moment of inertia, astronomers can infer
properties such as its mass and internal structure. Both the
Mass-Radius and Mass-Inertia relations are crucial tools for
understanding the properties and behavior of CSs, providing valuable
insights into the nature of these fascinating objects.

Here, we investigate the gravitational mass and radius with the
following condition $e^{-\beta}=1-\frac{2M}{R}$. The total mass is
obtained as
\begin{equation}\label{57a}
M=\frac{a^2R^3}{2(a^2R^2+(bR^2+1)^4)}.
\end{equation}
The relationship between the total mass (expressed in $M_{\odot}$)
and the radius (measured in $km$) is presented graphically in Figure
\textbf{9}. One notable finding from our graphical analysis is the
identification of the maximum mass exhibited by a star, which we
determine to be 1.3 solar mass. This particular observation
corresponds precisely with existing literature and is consistent
with the value provided in Table 1 of our study. Specifically, this
maximum mass was associated with the star identified as EXO
1785-248. Such compatibility between our findings and established
data reinforces the reliability and validity of our research
methodology and results.
\begin{figure}\center
\epsfig{file=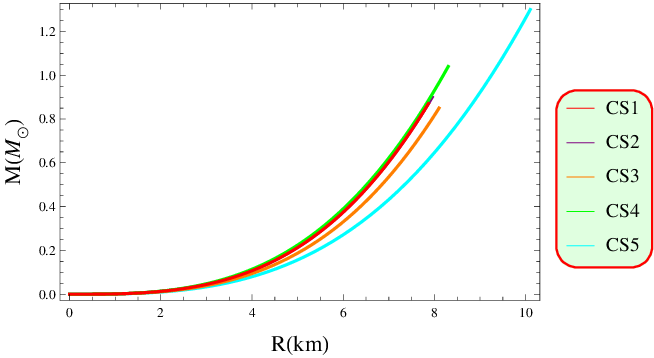,width=.5\linewidth}\caption{Graph of the total
mass corresponding to the total radius.}
\end{figure}
The study of moment of inertia is crucial in understanding the
rotational dynamics of compact objects. In the context of
astrophysics, the moment of inertia provides insight into how mass
is distributed in an object and how it affects its rotational
behavior. Understanding the relationship between mass and moment of
inertia is essential for modeling and predicting the behavior of
compact objects in astrophysical scenarios, providing valuable
insights into their structure and dynamics. The formula for the
moment of inertia as suggested by Bejger-Haensel \cite{44a}
\begin{equation}\label{57b}
I=\frac{2}{5}(1+M/M_{\odot})MR^2,
\end{equation}
allows us to relate the static properties of an object to its
rotational characteristics. The maximum mass of a uniformly slow
rotating configuration provides an approximate estimation of the
moment of inertia. This means that for a given mass and radius, the
moment of inertia reaches its maximum value when the object is
rotating at its slowest possible rate. Examining the nature of the
moment of inertia with respect to mass, it is evident that as mass
increases the moment of inertia also increases. This relationship is
depicted in Figure \textbf{10}. As mass increases, the more mass is
distributed farther away from the axis of rotation, resulting in a
larger moment of inertia. Eventually, the moment of inertia reaches
its maximum value for the given mass, indicating that further
increases in mass do not significantly affect the rotational
dynamics.
\begin{figure}\center
\epsfig{file=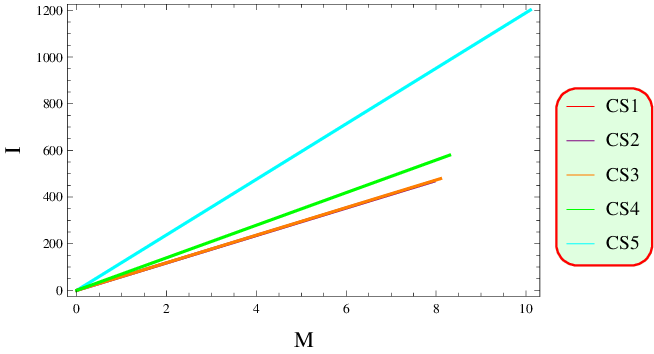,width=.5\linewidth}\caption{Plot of the moment
of inertia with respect to the mass.}
\end{figure}

\section{Equilibrium and Stability Analysis}

To Understand the structure and behavior of cosmic objects, it
requires a grasp of essential concepts such as equilibrium state and
stability analysis. An equilibrium state signifies a balance where
internal and external forces acting on cosmic structures are in
equilibrium. Stability is critical for assessing the consistency of
cosmic structures and involves exploring the conditions under which
these structures remain stable against different oscillation modes.
This analysis employs the concepts of \emph{sound speed} and
\emph{adiabatic index}. Sound speed reflects the speed at which
pressure waves propagate through a medium, while the adiabatic index
describes the relationship between pressure and density changes in
cosmic structures.

\subsection{Behavior of Various Forces}

Tolman-Opppenheimer equation is a fundamental equation which
describes the equilibrium structure of the static spherically
symmetric spacetime. It explains how the pressure and gravitational
forces of the star are balanced to maintain its equilibrium. This
equation describes the balance between the gravitational force
pulling the star inward and the pressure exerted by the degenerate
fermions or other constituents of the star pushing outward, leading
to an equilibrium configuration. In examining the equilibrium state
of stellar models, the TOV equation is crucial because it allows
astrophysicists to determine the relationship between the pressure,
density and mass distribution inside a star. The motivation for
exploring the TOV equation in the framework of CSs lies in
understanding the theoretical requirements and constraints for the
existence of stable CSs. By solving the TOV equation for various
configurations of CSs, we can obtain insights into the behavior of
these stars under extreme conditions. Specifically, we investigate
the relationship between crucial parameters for understanding their
stability and evolution. The TOV equation in the context of CSs
contributes to our understanding of the possibilities and
limitations of these fascinating constructs in the framework of
$f(\mathrm{Q},\mathrm{T})$ theory.

Tolman-Opppenheimer equation for anisotropic matter configuration is
\cite{45}
\begin{equation}\label{58}
\frac{M_{G}(r)e^\frac{\alpha-\beta}{2}}{r^{2}}(\varrho+p_{r})+
\frac{dp_{r}}{dr}-\frac{2}{r}(p_{t}-p_{r})=0,
\end{equation}
where
\begin{equation}\nonumber
M_{G}(r)=4\pi \int (\mathrm{T}^{t}_{t}-\mathrm{T}^{r}_{r} -
\mathrm{T}^{\theta}_{\theta}-\mathrm{T}^{\phi}_{\phi})r^{2}
e^{\frac{\alpha+\beta}{2}}dr.
\end{equation}
Solving this equation, we have
\begin{equation}\nonumber
M_{G}(r)=\frac{1}{2}r^{2}e^{\frac{\beta-\alpha}{2}}\alpha'.
\end{equation}
Substituting this value in Eq. (\ref{58}), we obtain
\begin{equation}\nonumber
\frac{1}{2}\alpha'(\varrho+p_{r})+
\frac{dp_{r}}{dr}-\frac{2}{r}(p_{t}-p_{r})=0.
\end{equation}
The resulting solution of this equation offers insights into the
star's internal structure, including details about its density
profile and pressure distribution. This analysis highlights the
impact of gravitational forces
($F_{g}=\frac{\mu'(\varrho+p_{r})}{2}$), hydrostatic forces
($F_{h}=\frac{dP_{r}}{dr}$) and anisotropic forces
($F_{a}=\frac{2(p_{r}-p_{t})}{r}$) on the system. Using
Eqs.(\ref{53})-(\ref{55}), we obtain
\begin{eqnarray}\nonumber
F_{g}&=&\bigg[4bdr(a+abr^{2})^{2}(2bd(1+br^{2})^{4}-2acb
(1+br^{2})(3br^2-1)+a^{2}d
\\\nonumber
&\times&(5br^{2}-1))\gamma\bigg]\bigg[(ad-2cb(1+br^{2}))^2
(a^{2}r^{2}+(1+br^{2})^{4})^{2}(1+\eta)\bigg]^{-1},
\\\nonumber
F_{a}&=&\bigg[2ar(a^{2}+4b(br^{2}+1)^{3})(a^{2}d-2acb(1+br^{2})
+2bd(1+br^{2})^{3})\gamma\bigg]
\\\nonumber
&\times&\bigg[(-ad+2cb(1+br^{2}))(a^{2}r^{2}+(1+br^{2})^{4})^{2}
(\eta+1)\bigg]^{-1},
\\\nonumber
F_{h}&=&\bigg[2ar\gamma(a^{7}d^{2}r^{2}(2\eta-3)-4a^{6}cbdr^{2}
(1+br^{2})(2\eta-3)+8a^{2}cb^{2}
\\\nonumber
&\times&d(1+br^{2})^{6}(9+br^{2}(18+br^{2}(9-23\eta)-41\eta)+32\eta)-8cb^{3}d
\\\nonumber
&\times&(1+br^{2})^{10}(-6-13\eta+br^{2}(17\eta-6))+2ab^{2}
(1+br^{2})^{8}((1+br^{2})
\\\nonumber
&\times&d^{2}(-6-23\eta+br^{2}(17\eta-6))+8c^{2}b(-3-14\eta
+br^{2}(26\eta-3)))
\\\nonumber
&+&4a^{4}bcd(1+br^{2})^{3}(3+14\eta+br^{2} (24-46\eta+(9+40\eta)
\\\nonumber
&\times&br^{2}))-2a^{3}b(1+br^{2})^{4}(d^{2}(1+br^{2})
(12+36\eta+br^{2}(12-59\eta+25
\\\nonumber
&\times&br^{2}\eta))+2c^{2}b(3+14\eta+br^2(18-44\eta
+br^{2}(15+38\eta))))-a^{5}
\\\nonumber
&\times&(1+br^{2})^{2}(4c^{2}b^{2}r^{2}(3-2\eta)+d^{2}
(3+14\eta+br^{2}(30-48\eta+br^{2}
\\\nonumber
&\times&(46\eta-9))))))\bigg]\bigg[3(ad-2cb(1+br^{2}))^{2}
(\eta+2\eta^2-1)
(a^{2}r^{2}+(1+br^{2})^{4})^{3}\bigg]^{-1}.
\end{eqnarray}
Figure \textbf{11} shows that our considered CSs are in equilibrium
state as the total effect of $F_{g}$, $F_{h}$ and $F_{a}$ is zero.
\begin{figure}\center
\epsfig{file=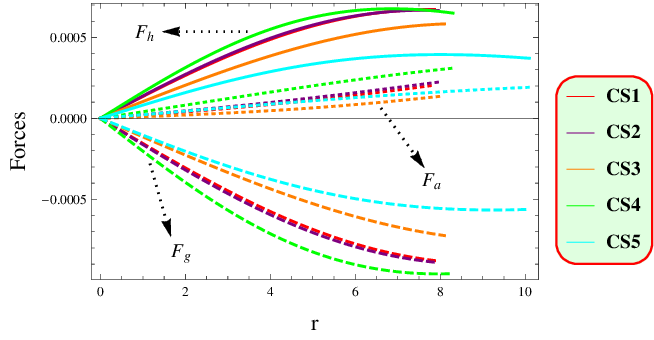,width=.5\linewidth}\caption{Plot of TOV
equation versus radial coordinate.}
\end{figure}

\subsection{Casuality Condition}

The stability analysis of CSs can be examined through the causality
condition, ensuring that signals or information do not propagate
faster than the speed of light. This criterion dictates that the
radial and tangential components of sound speed, denoted as
$(u_{r}=\frac{dp_{r}}{d\varrho})$ and $(u_{t}
=\frac{dp_{t}}{d\varrho})$ must be confined in the interval [0,1]
for structures to be considered stable \cite{47}. The expressions
for the components of sound speed in the context of
$f(\mathrm{Q},\mathrm{T})$ are provided in Appendix \textbf{C}.
Figure \textbf{12} visually demonstrates that the static spherically
symmetric solutions adhere to the required constraints, affirming
their stability.
\begin{figure}
\epsfig{file=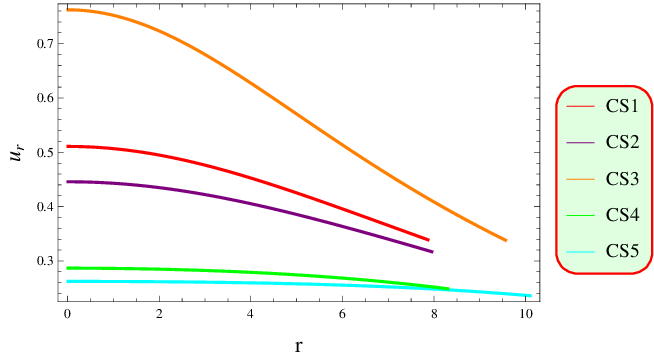,width=.5\linewidth}
\epsfig{file=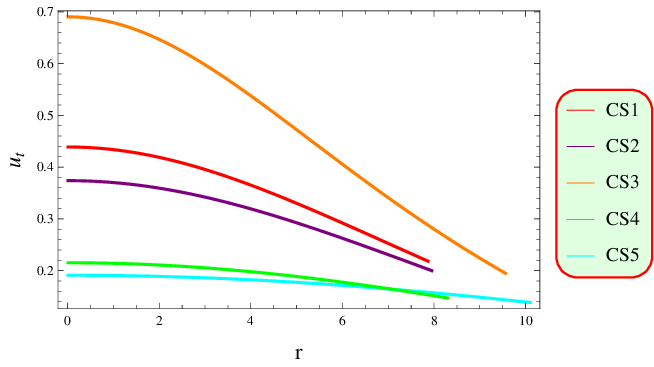,width=.5\linewidth}\caption{Plots of sound
speed versus radial coordinate.}
\end{figure}

\subsection{Herrera Cracking Approach}

The Herrera cracking technique serves as a mathematical tool for
examining the stability of solutions. In assessing the stability of
CSs, one can evaluate the cracking condition, expressed as $(0 \leq
|u_{t} - u_{r}| \leq 1)$ \cite{47}. The violation of this cracking
condition implies instability, leading to the collapse of CSs.
Conversely, the satisfaction of the cracking condition indicates
stability, allowing the CSs to persist over an extended period.
Figure \textbf{13} confirms the stability of the considered CSs, as
they remain in the prescribed limits.
\begin{figure}\center
\epsfig{file=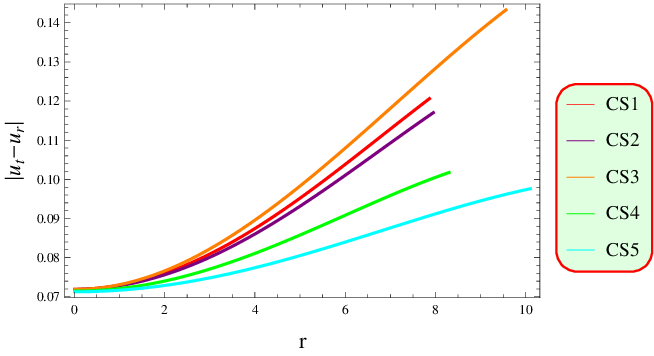,width=.5\linewidth}\caption{Behavior of Herrera
cracking approach.}
\end{figure}

\subsection{Adiabatic Index}

This technique provides information about how matter responds to
changes in pressure and density and can be used to determine whether
CSs are stable. The adiabatic index is defined as
\begin{eqnarray}\nonumber
\Gamma_{r}=\frac{\varrho+p_{r}}{p_{r}} u_{r},\quad
\Gamma_{t}=\frac{\varrho+p_{t}}{p_{t}} u_{t},
\end{eqnarray}
where $\Gamma_{r}$ and $\Gamma_{t}$ are the radial and tangential
components of the adiabatic index. The expressions for $\Gamma_{r}$
and $\Gamma_{t}$ using Eqs.(\ref{53})-(\ref{55}) are provided in
Appendix \textbf{D}. If the value of $\Gamma$ is less than 4/3, then
CSs are stable, otherwise CSs are unstable and will collapse
\cite{48}. Figure \textbf{14} shows that our system is stable in the
presence of correction terms, as it satisfies the required limit.
\begin{figure}
\epsfig{file=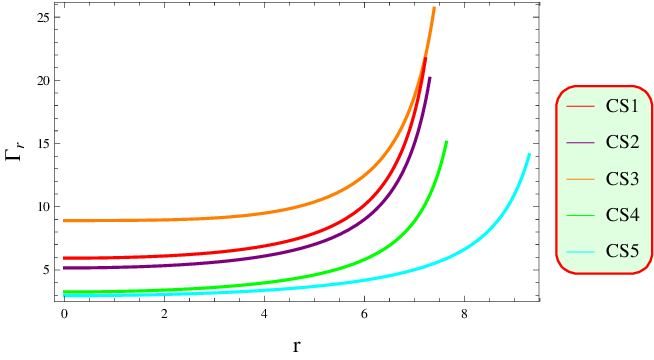,width=.5\linewidth}
\epsfig{file=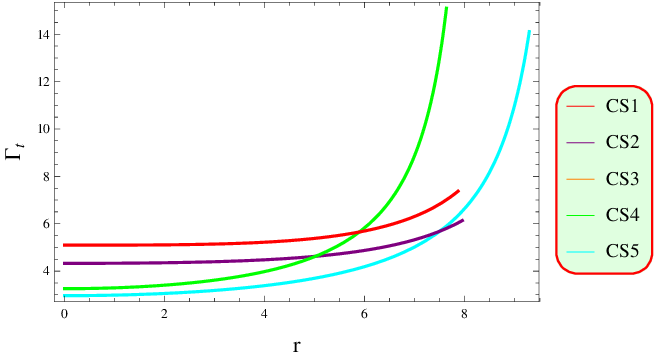,width=.5\linewidth}\caption{Behavior of
adiabatic index versus radial coordinate.}
\end{figure}

\section{Conclusions}

Modified gravity theories, including $f(\mathrm{Q},\mathrm{T})$
gravity have gained significant attention in recent years as
alternative explanations to the EGTR. This theory proposes
modification to the fundamental equations of gravity, aiming to
address various unresolved questions and challenges in cosmology and
astrophysics. One particularly important avenue for testing and
constraining this modified gravity theory is through astrophysical
observations. Stars, being fundamental objects in the universe which
provide an excellent laboratory for investigating the effects of
modified gravity. Observations of stars can help us to discern the
differences in the behavior of gravity predicted by
$f(\mathrm{Q},\mathrm{T})$ theory as compared to EGTR. Stellar
properties such as mass, radius, luminosity and evolutionary stages
can be affected by modification to the gravitational force law,
offering opportunities to test and differentiate between
$f(\mathrm{Q},\mathrm{T})$ theory and EGTR. This modified theory
often introduces additional free parameters that characterize
deviations from EGTR. By studying stars in this theory, we can place
constraints on these parameters by comparing theoretical predictions
with observed properties of stars, providing valuable insights into
the viability of this theory. The study of stars in this modified
gravity contributes to our understanding of the fundamental nature
of gravity itself. It helps to refine our knowledge of the laws that
govern the behavior of matter and energy in the universe and opens
avenues for exploring new phenomena that might not be present in
EGTR.

Over the past two decades, the study of CSs has captivated the field
of theoretical physics. These cosmic objects pose significant
challenges to our comprehension of fundamental physical and present
unique opportunities to investigate extreme conditions in the
theoretical framework. Our research focuses on examining compact
stellar structures in the framework of modified
$f(\mathrm{Q},\mathrm{T})$ theory, aiming to delve into the
mysteries of the universe. This modified theory holds promise in
elucidating phenomena associated with the dark universe. Through the
exploration of CSs in this theoretical framework, we have unveiled
novel gravitational interactions at both galactic and cosmological
scales, providing insight into the nature of these components and
their impact on stellar structures. The gravity in compact stellar
objects reaches its most extreme limits. Therefore, studying these
objects in this theory allows us to probe gravity behavior under
conditions of high curvature and density. By scrutinizing the
behavior of these extreme celestial objects, we have gained valuable
insights into the characteristics of CSs, advancing our
comprehension of fundamental interactions in the universe. This
research stands to revolutionize our understanding of gravity,
opening pathways to a deep understanding of the cosmos and the
forces shaping it.

The proposed model of $f(\mathrm{Q},\mathrm{T})$ theory in the
context of CSs have significant implications for the behavior of
matter and geometry. In CSs such as neutron stars or quark stars,
the high density and extreme conditions lead to unique physical
phenomena. The inclusion of $f(\mathrm{Q},\mathrm{T})$ terms in the
gravitational action affect the behavior of matter in these stars.
For instance, it can influence the EOS describing the relationship
between pressure and density inside the star. This affects the
structure and composition of the star. The modified gravitational
action lead to deviations from the predictions of EGTR in terms of
the spacetime geometry around CSs. The curvature of spacetime as
described by the metric tensor is influenced by the additional
modified terms. This can result in modifications to the
gravitational field equations, affecting the metric components and
curvature profiles around the star. Observationally, the effects of
this $f(\mathrm{Q},\mathrm{T})$ model on CSs can manifest in various
ways. This model offers a promising framework to explore
modifications to gravity in the context of CSs. These include
theoretical consistency, constraints from observational data and
distinguishing the effects of $f(\mathrm{Q},\mathrm{T})$
modifications from other astrophysical phenomena. Thus, the minimal
model of $f(\mathrm{Q},\mathrm{T})$ theory can significantly impact
the behavior of matter and geometry in CSs, potentially leading to
observable consequences that may provide insights into the nature of
gravity and the properties of dense matter in extreme astrophysical
environments.

In our study, we have used the matching conditions at the boundary
between the interior and exterior spacetimes of CSs to determine the
values of the unknown constants. Specifically, we have used boundary
conditions derived from the junction conditions, ensuring continuity
of the metric potentials and their derivatives across the boundary.
By solving the resulting system of equations, we have obtained the
values of the unknown constants that satisfy these matching
conditions. Regarding the implications of these values for the
behavior and properties of CSs, the determined constants play a
crucial role in characterizing the spacetime geometry and matter
distribution in the stellar interior. The specific values of these
constants directly impact the structural and dynamical properties of
CSs, including their stability, maximum mass, compactness and
overall gravitational behavior. We acknowledge the importance of
elucidating this aspect of our methodology and its significance for
understanding the behavior of CSs.

We have used the TOV equation to analyze the equilibrium state of
the stellar models under consideration. The results obtained from
our analysis provide valuable insights into the stability of CSs.
Our study contributes to the understanding of the behavior of CSs
and their stability under extreme conditions. We believe that the
insights gained from our analysis have important implications for
astrophysical phenomena. By incorporating modified gravitational
theory and the TOV equation, we have explored implications for
theoretical physics and the study of spacetime geometry beyond
astrophysical scenarios.

The key findings are summarized as follows.
\begin{itemize}
\item
The metric functions (Figure \textbf{1}) are consistent and
non-singular, ensuring that the spacetime is smooth and free from
singularities.
\item
The behavior of matter contents (Figure \textbf{2}) is positive and
maximum at the center of the CSs, implying that it has a stable core
and decreases towards the boundary, making the CSs physically
viable. Moreover, the radial pressure vanishes at the surface
boundary ($r=\mathcal{R}$), i.e., $p_{r}(r=\mathcal{R})=0$.
\item
The gradient of matter contents (Figure \textbf{3}) vanishes at the
center of CSs and then shows negative behavior, presenting a dense
picture of the considered CSs.
\item
The pressure components are equal at center, which demonstrates that
the anisotropy vanishing at the center of CSs. The positive behavior
of anisotropy (Figure \textbf{4}) indicates that pressure is
directed outward, which is necessary for CSs.
\item
All energy constraints (Figure \textbf{5}) are positive, ensuring
the presence of ordinary matter in the interior of CSs, which is
necessary to obtain viable CSs.
\item
The EoS parameters (Figure \textbf{6}) satisfy the required
constraints, i.e., $0 < \omega_{r},\omega_{t} < 1$, which shows the
viability of the considered model.
\item
We have found that the mass function (Figure \textbf{7}) is regular
at the center of CSs and shows a monotonically increasing behavior
as the radial coordinate increases.
\item
The compactness factor is less than 4/9, and the redshift is less
than 5.2, leading to viable CSs structures (Figure \textbf{8}).
\item
The graphical behavior of $M-R$ and $M-I$ relations are given in
Figures \textbf{9} and \textbf{10}
\item
Tolman-Opppenheimer equation (Figure \textbf{11}) shows that all
forces (gravitational, hydrostatic, and anisotropic) have a null
impact for all proposed CSs. This suggests that the CSs are in an
equilibrium state.
\item
The stability limits i.e. causality condition, Herrera cracking
approach and adiabatic index are satisfied ensuring the existence of
physically stable CSs (Figures \textbf{12}-\textbf{14}).
\end{itemize}
Salako et al \cite{48a} studied the existence of strange stars in
the background of $f(\mathbb{T}, \mathrm{T})$ gravity, where
$\mathbb{T}$ is the torsion tensor. They examined the compact
spherical structure of only LMC X-4 strange star candidate using
conformal Killing vectors and ensures their stability through
several physical parameters. Das et al \cite{48b} obtained a new
class of interior solutions to the Einstein field equations for an
anisotropic matter distribution using a linear EoS. They verified
the physical acceptability of the solutions by the current estimated
data of pulsar 4U-160852. In contrast, our study diverges by
adopting the $f(\mathrm{Q},\mathrm{T})$ theory, a different
theoretical framework that allows for a comprehensive exploration of
the effects of modified terms on the viability and stability of CSs.
Using distinct methodologies and techniques, we have analyzed
several anisotropic CSs, discerning their viability and stability
under the influence of modified gravitational dynamics. Our findings
reveal the feasibility of the proposed CSs even in the presence of
modified terms, underscoring the robustness of our theoretical
framework. By expanding the scope of our investigation to encompass
a broader spectrum of CSs, we contribute to the ongoing discourse on
the behavior of CSs in alternative gravitational theory, thus
enriching our understanding of astrophysical phenomena across
diverse theoretical landscapes.

We have checked whether our system remains in stable state or not in
the presence of $f(\mathrm{Q},\mathrm{T})$ terms. It is found that
in the presence of modified terms our system remains in a stable
state. Our investigation into the physical aspects has revealed a
dense profile of CSs. We have analyzed various essential physical
parameters such as metric potentials, effective matter variables,
energy conditions, mass, compactness, redshift function, TOV
equation, sound speed and adiabatic index, which characterize the
stellar system. It is worthwhile to mention here that all these
parameters meet the necessary conditions, indicating the presence of
viable and stable CSs in this modified framework. Furthermore, the
chosen factors for assessing the feasibility and stability of the
solution have satisfied their specified limits. Notably, we observed
that all parameters reach their maximum values when compared to EGTR
and other modified gravitational theories. In the realm of
$f(\textrm{R})$ theory, the results indicate the instability of the
Her X-1 CS associated with the second gravity model due to the
limited range satisfied by the physical quantities \cite{51}.
Furthermore, in the framework of $f(\textrm{R},\textrm{T}^{2})$
theory, it is found that CSs are neither physically viable nor
stable at the center \cite{52}. In the light of these findings, it
can be concluded that all considered CSs exhibit both physical
viability and stability at the center in this modified theory.
Consequently, our results suggest that viable and stable CSs can
exist in this modified theory. Therefore, we conclude that the
solutions we have obtained are physically valid, providing stable
and viable structures for anisotropic CSs.

\section*{Appendix A: Non-Metricity Scalar}
\renewcommand{\theequation}{A\arabic{equation}}
\setcounter{equation}{0}

According to Eqs.(\ref{18}) and (\ref{19}), we have
\begin{eqnarray}\nonumber
\mathrm{Q} &\equiv& -\mathrm{g}^{\mu\nu}
(\mathrm{L}^{\lambda}_{\xi\mu}\mathrm{L}^{\xi}_{\nu\lambda} -
\mathrm{L}^{\lambda}_{\xi\lambda}\mathrm{L}^{\xi}_{\mu\nu}),
\\\nonumber
\mathrm{L}^{\lambda}_{\xi\mu}&=&-\frac{1}{2}
\mathrm{g}^{\lambda\varsigma}(\mathrm{Q}_{\mu\xi\varsigma}
+\mathrm{Q}_{\xi\varsigma\mu}-\mathrm{Q}_{\varsigma\mu\xi}),
\\\nonumber
\mathrm{L}^{\xi}_{\nu\lambda}&=&-\frac{1}{2}
\mathrm{g}^{\xi\varsigma}(\mathrm{Q}_{\lambda\nu\varsigma}
+\mathrm{Q}_{\nu\varsigma\lambda}-\mathrm{Q}_{\varsigma\lambda\nu}),
\\\nonumber
\mathrm{L}^{\lambda}_{\xi\mu} &=& -\frac{1}{2}
\mathrm{g}^{\lambda\varsigma}(\mathrm{Q}_{\lambda\xi\varsigma}
+\mathrm{Q}_{\xi\varsigma\lambda}-\mathrm{Q}_{\varsigma\lambda\xi}),
\\\nonumber
&=&-\frac{1}{2}(\bar{\mathrm{Q}}_{\xi}
+\mathrm{Q}_{\xi}-\bar{\mathrm{Q}}_{\xi})=-\frac{1}{2}
\mathrm{Q}_{\xi},
\\\nonumber
\mathrm{L}^{\xi}_{\mu\nu}&=&
-\frac{1}{2}\mathrm{g}^{\xi\varsigma}(\mathrm{Q}_{\nu\mu\varsigma}
+\mathrm{Q}_{\mu\varsigma\nu}-\mathrm{Q}_{\varsigma\nu\mu}).
\end{eqnarray}
Thus, we have
\begin{eqnarray}\nonumber
-\mathrm{g}^{\mu\nu}\mathrm{L}^{\lambda}_{\xi\mu}
\mathrm{L}^{\xi}_{\nu\lambda}&=&
-\frac{1}{4}\mathrm{g}^{\mu\nu}\mathrm{g}^{\lambda\varsigma}
\mathrm{g}^{\xi\varsigma}
(\mathrm{Q}_{\mu\xi\varsigma}+\mathrm{Q}_{\xi\varsigma\mu}
-\mathrm{Q}_{\varsigma\mu\xi})
\\\nonumber
&\times&(\mathrm{Q}_{\lambda\nu\varsigma}+\mathrm{Q}_{\nu\varsigma\lambda}
-\mathrm{Q}_{\varsigma\lambda\nu}),
\\\nonumber
&=&-\frac{1}{4}(\mathrm{Q}^{\nu\varsigma\lambda}+\mathrm{Q}^{\varsigma\lambda\nu}
-\mathrm{Q}^{\lambda\nu\varsigma})
\\\nonumber
&\times&(\mathrm{Q}_{\lambda\nu\varsigma}+\mathrm{Q}_{\nu\varsigma\lambda}
-\mathrm{Q}_{\varsigma\lambda\nu}),
\\\nonumber
&=&-\frac{1}{4}(2\mathrm{Q}^{\nu\varsigma\lambda}\mathrm{Q}_{\varsigma\lambda\nu}
- \mathrm{Q}^{\nu\varsigma\lambda}\mathrm{Q}_{\nu\varsigma\lambda}),
\\\nonumber
\mathrm{g}^{\mu\nu}\mathrm{L}^{\lambda}_{\xi\lambda}\mathrm{L}^{\xi}_{\mu\nu}&=&
\frac{1}{4}\mathrm{g}^{\mu\nu}\mathrm{g}^{\xi\varsigma}\mathrm{Q}_{\varsigma}
(\mathrm{Q}_{\nu\mu\varsigma}+\mathrm{Q}_{\mu\varsigma\nu}
-\mathrm{Q}_{\varsigma\nu\mu}),
\\\nonumber
&=&\frac{1}{4}\mathrm{Q}^{\varsigma}(2\bar{\mathrm{Q}_{\varsigma}}-
\mathrm{Q}_{\varsigma}),
\\\nonumber
\mathrm{Q}&=&
-\frac{1}{4}(\mathrm{Q}^{\lambda\nu\varsigma}\mathrm{Q}_{\lambda\nu\varsigma}
+2\mathrm{Q}^{\lambda\nu\varsigma\lambda}\mathrm{Q}_{\varsigma\lambda\nu}
\\\nonumber
&-&2\mathrm{Q}^{\varsigma}\bar{\mathrm{Q}_{\varsigma}}+\mathrm{Q}^{\varsigma}\mathrm{Q}_{\varsigma}).
\end{eqnarray}
According to Eq.(\ref{24}), we obtain
\begin{eqnarray}\nonumber
\mathrm{P}^{\lambda\mu\nu}&=&\frac{1}{4}[-\mathrm{Q}^{\lambda\mu\nu}
+\mathrm{Q}^{\mu\lambda\nu}+\mathrm{Q}^{\nu\lambda\mu}
+\mathrm{Q}^{\lambda}\mathrm{g}^{\mu\nu}
\\\nonumber
&-&\bar{\mathrm{Q}^{\lambda}}\mathrm{g}^{\mu\nu}-\frac{1}{2}
(\mathrm{g}^{\lambda\mu}
\mathrm{Q}^{\nu}+\mathrm{g}^{\lambda\nu}\mathrm{Q}^{\mu})],
\\\nonumber
-\mathrm{Q}_{\lambda\mu\nu}\mathrm{P}^{\lambda\mu\nu} &=&
-\frac{1}{4}[-\mathrm{Q}_{\lambda\mu\nu}\mathrm{Q}^{\lambda\mu\nu}
\\\nonumber
&+&\mathrm{Q}_{\lambda\mu\nu}\mathrm{Q}^{\mu\lambda\nu}
+\mathrm{Q}^{\nu\lambda\mu}
\mathrm{Q}_{\lambda\mu\nu}+\mathrm{Q}_{\lambda\mu\nu}
\mathrm{Q}^{\lambda}\mathrm{g}^{\mu\nu}
\\\nonumber
&-&\mathrm{Q}_{\lambda\mu\nu}\bar{\mathrm{Q}^{\lambda}}
\mathrm{g}^{\mu\nu}
-\frac{1}{2}\mathrm{Q}_{\lambda\mu\nu}(\mathrm{g}^{\lambda\mu}
\mathrm{Q}^{\nu} +\mathrm{g}^{\lambda\nu}\mathrm{Q}^{\mu})],
\\\nonumber
&=&
-\frac{1}{4}(-\mathrm{Q}_{\lambda\mu\nu}\mathrm{Q}^{\lambda\mu\nu}
+2\mathrm{Q}_{\lambda\mu\nu}\mathrm{Q}^{\mu\lambda\nu}+\mathrm{Q}^{\lambda}
\mathrm{Q}_{\lambda}-2\tilde{\mathrm{Q}^{\lambda}}\mathrm{Q}_{\lambda}),
\\\nonumber
&=& \mathrm{Q}.
\end{eqnarray}

\section*{Appendix B: Variation of Non-Metricity Scalar}
\renewcommand{\theequation}{B\arabic{equation}}
\setcounter{equation}{0}

All the non-metricity tensors are given as
\begin{eqnarray}\nonumber
\mathrm{Q}_{\lambda\mu\nu}&=&\nabla_{\lambda}\mathrm{g}_{\mu\nu},
\\\nonumber
\mathrm{Q}^{\lambda}~_{\mu\nu}&=&\mathrm{g}^{\lambda\xi}
\mathrm{Q}_{\xi\mu\nu}
=\mathrm{g}^{\lambda\xi}\nabla_{\xi}\mathrm{g}_{\mu\nu}
=\nabla^{\lambda}\mathrm{g}_{\mu\nu},
\\\nonumber
\mathrm{Q}_{\lambda~~\nu}^{~~\mu}&=&\mathrm{g}^{\mu\varsigma}\mathrm{Q}_{\lambda\varsigma\nu}
=\mathrm{g}^{\mu\varsigma}\nabla_{\lambda}\mathrm{g}_{\varsigma\nu}
=-\mathrm{g}_{\mu\varsigma}\nabla_{\lambda}\mathrm{g}^{\mu\varsigma},
\\\nonumber
\mathrm{Q}_{\lambda\mu}^{~~\nu} &=&
\mathrm{g}^{\nu\varsigma}\mathrm{Q}_{\lambda\mu\varsigma}
=\mathrm{g}^{\nu\varsigma}\nabla_{\lambda}\mathrm{g}_{\mu\varsigma}
=-\mathrm{g}_{\mu\varsigma}\nabla_{\lambda}\mathrm{g}^{\nu\varsigma},
\\\nonumber
\mathrm{Q}^{\lambda\mu}_{~~\nu}&=&
\mathrm{g}^{\mu\varsigma}\mathrm{g}^{\lambda\xi}\nabla_{\xi}\mathrm{g}
_{\varsigma\nu}
=\mathrm{g}^{\mu\varsigma}\nabla^{\lambda}\mathrm{g}_{\nu\varsigma}
=-\mathrm{g}_{\varsigma\nu}\nabla^{\lambda}\mathrm{g}^{\mu\varsigma},
\\\nonumber
\mathrm{Q}^ {\lambda~~\nu} _{~\mu} &=&
\mathrm{g}^{\nu\varsigma}\mathrm{g}^{\lambda\xi}\nabla_{\xi}\mathrm{g}
_{\mu\varsigma}
=\mathrm{g}^{\nu\varsigma}\nabla^{\lambda}\mathrm{g}_{\mu\varsigma}
=-\mathrm{g}_{\mu\varsigma}\nabla^{\lambda}\mathrm{g}^{\nu\varsigma},
\\\nonumber
\mathrm{Q}_{\lambda}^{~~\mu\nu}&=&
\mathrm{g}^{\mu\varsigma}\mathrm{g}^{\nu\xi}\nabla_{\lambda}\mathrm{g}
_{\varsigma\xi}
=-\mathrm{g}^{\mu\varsigma}\mathrm{g}_{\varsigma\xi}\nabla_{\lambda}
\mathrm{g}^{\nu\varsigma} =-\nabla_{\lambda}\mathrm{g}^{\mu\nu}.
\end{eqnarray}
By using Eq.(\ref{25}), we have
\begin{eqnarray}\nonumber
\delta \mathrm{Q} &=&-\frac{1}{4}
\delta(-\mathrm{Q}^{\lambda\nu\varsigma}
\mathrm{Q}_{\lambda\nu\varsigma}+2\mathrm{Q}^{\lambda\nu\varsigma}
\mathrm{Q}_{\varsigma\lambda\nu}-2\mathrm{Q}^{\varsigma}
\bar{\mathrm{Q}_{\varsigma}}+\mathrm{Q}^{\varsigma}\mathrm{Q}_{\varsigma}),
\\\nonumber
&=&-\frac{1}{4}(-\delta \mathrm{Q}^{\lambda\nu\varsigma}
\mathrm{Q}_{\lambda\nu\varsigma} -
\mathrm{Q}^{\lambda\nu\varsigma}\delta
\mathrm{Q}_{\lambda\nu\varsigma} + 2\delta
\mathrm{Q}_{\lambda\nu\varsigma}\mathrm{Q}^{\varsigma\lambda\nu}
\\\nonumber
&+& 2 \mathrm{Q}^{\lambda\nu\varsigma}\delta
\mathrm{Q}_{\varsigma\lambda\nu}-2\delta
\mathrm{Q}^{\varsigma}\bar{\mathrm{Q}_{\varsigma}}+\delta
\mathrm{Q}^{\varsigma}\mathrm{Q}_{\varsigma}-2
\mathrm{Q}^{\varsigma}\delta \bar {\mathrm{Q}_{\varsigma}} +
\mathrm{Q}^{\varsigma}\delta \mathrm{Q}_{\varsigma}),
\\\nonumber
&=&-\frac{1}{4}[\mathrm{Q}_{\lambda\nu\varsigma}\nabla
^{\lambda}\delta
\mathrm{g}^{\nu\varsigma}-\mathrm{Q}^{\lambda\nu\varsigma}
\nabla_{\lambda}\delta
\mathrm{g}_{\nu\varsigma}-2\mathrm{Q}_{\varsigma\lambda\nu}
\nabla^{\lambda}\delta \mathrm{g}^{\nu\varsigma}
\\\nonumber
&+&2\mathrm{Q}^{\lambda\nu\varsigma}\nabla_{\varsigma}\delta
\mathrm{g}_{\lambda\nu}+
2\bar{\mathrm{Q}_{\varsigma}}\mathrm{g}^{\mu\nu}\nabla
^{\varsigma}\delta
\mathrm{g}_{\mu\nu}+2\mathrm{Q}^{\varsigma}\nabla^{\xi}\delta
\mathrm{g}_{\varsigma\xi}
\\\nonumber
&+&2\bar{\mathrm{Q}_{\varsigma}}
\mathrm{g}_{\mu\nu}\nabla^{\varsigma}\delta
\mathrm{g}^{\mu\nu}-\mathrm{Q}_{\varsigma}\nabla^{\xi}\mathrm{g}
^{\mu\nu}\delta
\mathrm{g}_{\mu\nu}-\mathrm{Q}_{\varsigma}\mathrm{g}_{\mu\nu}\nabla^{\varsigma}\delta
\mathrm{g}^{\mu\nu}
\\\nonumber
&-&\mathrm{Q}_{\varsigma} \mathrm{g}^{\mu\nu} \nabla_{\varsigma}
\delta \mathrm{g}_{\mu\nu}
-\mathrm{Q}^{\varsigma}\mathrm{g}_{\mu\nu}\nabla_{\varsigma}\delta
\mathrm{g}_{\mu\nu}].
\end{eqnarray}
We use the following relations to simplify the above equation
\begin{eqnarray}\nonumber
\delta \mathrm{g}_{\mu\nu}&=&-\mathrm{g}_{\mu\lambda} \delta
\mathrm{g}^{\lambda\xi}\mathrm{g}_{\xi\nu}-\mathrm{Q}^
{\lambda\nu\varsigma} \nabla_{\lambda}\delta
\mathrm{g}_{\nu\varsigma},
\\\nonumber
\delta \mathrm{g}_{\nu\varsigma}&=&-\mathrm{Q}^{\lambda\nu\varsigma}
\nabla_{\lambda}(-\mathrm{g}_{\nu\mu}\delta
\mathrm{g}^{\mu\xi}\mathrm{g}_{\xi\varsigma}),
\\\nonumber
&=&2\mathrm{Q}_{~~\varsigma}^{\lambda\nu}\mathrm{Q}_{\lambda\nu\mu}\delta
\mathrm{g}^{\mu\varsigma} +
\mathrm{Q}_{\lambda\xi\varsigma}\nabla^{\lambda}\mathrm{g}^{\mu\varsigma}
\\\nonumber
&=&2\mathrm{Q}_{~~\nu}^{\lambda\xi}\mathrm{Q}_{\lambda\xi\nu}\delta
\mathrm{g}^{\mu\nu}+\mathrm{Q}_{\lambda\nu\varsigma}\nabla^{\lambda}
\mathrm{g}^{\nu\varsigma},
\\\nonumber
2\mathrm{Q}^{\lambda\nu\varsigma}\nabla_{\varsigma}\delta
\mathrm{g}_{\lambda\nu}&=&
-4\mathrm{Q}_{\mu}^{~\xi\varsigma}\mathrm{Q}_{\varsigma\xi\nu}
\delta \mathrm{g}^{\mu\nu}-2
\mathrm{Q}_{\nu\varsigma\lambda}\nabla^{\lambda}\delta
\mathrm{g}^{\nu\varsigma},
\\\nonumber
-2\mathrm{Q}^{\varsigma} \nabla^{\xi} \delta
\mathrm{g}_{\varsigma\xi}&=&2\mathrm{Q}^{\lambda}
\mathrm{Q}_{\nu\lambda\mu} \delta \mathrm{g}^{\mu\nu}+
2\mathrm{Q}_{\mu}\bar{\mathrm{Q}_{\nu}} \delta \mathrm{g}^{\mu\nu}
\\\nonumber
&+&2\mathrm{Q}_{\nu}\mathrm{g}_{\lambda\varsigma}\nabla^{\lambda}
\mathrm{g}^{\nu\varsigma}.
\end{eqnarray}
Thus, we have
\begin{equation}\nonumber
\delta\mathrm{Q}=2\mathrm{P}_{\lambda\nu\varsigma}\nabla^{\lambda}\delta
\mathrm{g}^{\nu\varsigma}-(\mathrm{P}_{\mu\lambda\xi}\mathrm{Q}_{\nu}
^{\lambda\xi}-2 \mathrm{P}_{\lambda\xi\nu}\mathrm{Q}^{\lambda\xi}
_{\nu})\delta \mathrm{g}^{\mu\nu},
\end{equation}
where
\begin{eqnarray}\nonumber
2\mathrm{P}_{\lambda\nu\varsigma}&=&-\frac{1}{4}[2\mathrm{Q}
_{\lambda\nu\varsigma}-2\mathrm{Q}_{\varsigma\lambda\nu}
-2\mathrm{Q}_{\nu\varsigma\lambda}
\\\nonumber
&+&2(\bar{\mathrm{Q}}_{\lambda}-\mathrm{Q}_{\lambda})
\mathrm{g}_{\nu\varsigma}+2\mathrm{Q}_{\nu}\mathrm{g}_{\lambda\xi}],
\\\nonumber
4(\mathrm{P}_{\mu\lambda\xi}\mathrm{Q}_{\nu}^{~~\lambda\xi}-2
\mathrm{P}_{\lambda\xi\nu}\mathrm{Q}^{\lambda\xi}
_{~~\nu})&=&2\mathrm{Q}^{\lambda\xi}
_{~~\nu}\mathrm{Q}_{\lambda\xi\mu}-4\mathrm{Q}_{\mu}
^{~~\lambda\xi}\mathrm{Q}_{\xi\lambda\nu}+2\mathrm{Q}_{\lambda\mu\nu}
\bar{\mathrm{Q}}^{\lambda}
\\\nonumber
&-&\mathrm{Q}^{\lambda}\mathrm{Q}_{\lambda\mu\nu}
+2\mathrm{Q}^{\lambda}\mathrm{Q}_{\nu\lambda\mu}+2\mathrm{Q}
_{\mu}\bar{\mathrm{Q}_{\nu}}.
\end{eqnarray}

\section*{Appendix C}
\renewcommand{\theequation}{C\arabic{equation}}
\setcounter{equation}{0}

The radial and tangential components of sound speed are given by
\begin{eqnarray}\nonumber
u_{sr}&=&\bigg[a^{7}d^{2}r^{2}(3-2\eta)+4a^{6}cbdr^{2}
(1+br^{2})(2\eta-3)+8cb^{3}(1+br^{2})^{10}d(-6
\\\nonumber
&-&13\eta+br^{2}(17\eta-6))+8a^{2}cb^{2}d(1+br^{2})^{6}
\big(-9-32\eta+br^{2}(-18+41\eta
\\\nonumber
&+&br^{2}(23\eta-9))\big)-2ab^{2}(1+br^{2})^{8}(d^{2}(1+br^{2})
(-23\eta-6+br^{2}(17\eta-6))
\\\nonumber
&+&8c^{2}b(-3-14\eta+br^{2}(26\eta-3)))-4a^{4}cbd(1+br^{2})
^{3}(3+14\eta+br^{2}(24
\\\nonumber
&-&46\eta+br^{2}(9+40\eta)))+(1+br^{2})^{4}2a^{3}b(d^{2}(1+br^{2})
(12+36\eta+br^{2}(12
\\\nonumber
&-&59\eta+25br^{2}\eta))+2c^{2}b(14\eta+3+br^{2}(18-44\eta+br^{2}
(15+38\eta))))
\\\nonumber
&+&a^{5}(1+br^{2})^{2}\big(4c^{2}(3-2\eta)b^{2}r^{2}+d^{2}(3+14\eta
+br^{2}(30-48\eta+br^{2}
\\\nonumber
&\times&(46\eta-9)))\big)\bigg]\bigg[-40cb^{3}d(br^{2}-5)(1+br^{2})
^{10}\eta+a^{7}d^{2}r^{2}(2\eta-3)
\\\nonumber
&-&4a^{6}cbr^{2}(1+br^{2})d(-3+2\eta)+10ab^{2}(1+br^{2})^{8}
\big(d^{2}(br^{2}-7)(1+br^{2})\eta
\\\nonumber
&-&8c^{2}b(br^2-1)(2\eta-3)\big)+40a^{2}cb^{2}d(1+br^{2})^{6}(6-2
\eta-br^{2}\eta+b^{2}r^{4}
\\\nonumber
&\times&(11\eta-6))-4a^{4}cbd(1+br^{2})^{3}(5(-3+2\eta)+br^{2}
(6-14\eta+br^{2}
\\\nonumber
&\times&(44\eta-51)))-2a^{3}b(1+br^{2})^{4}\big(-2c^{2}b(5+(-2
+17br^{2}))br^{2}(2\eta-3)
\\\nonumber
&+&5d^{2}(1+br^{2})(6+br^{2}(-7\eta+br^{2}(17\eta-6)))\big)+a^{5}
(br^{2}+1+b(-7\eta
\\\nonumber
&+&br^{2}(17\eta-6)))+a^{5}(1+br^{2})^{2}(4(2\eta-3)c^{2}b^{2}
r^{2}+d^{2}(5(2\eta-3)+br^{2}
\\\nonumber
&\times&(6-24\eta+br^{2}(74\eta-51))))\bigg]^{-1},
\\\nonumber
u_{st}&=&\bigg[2(2a^{7}d^{2}r^{2}\eta-8a^{6}cbdr^{2}(1+br^{2})
\eta-4cb^{3}d(1+br^{2})^{10}((7\eta-6)r^{2}b
\\\nonumber
&+&12+\eta)+ab^{2}(1+br^{2})^{8}\big[-8c^{2}b(-6-8\eta+br^{2}(9+2\eta))
+d^{2}(1+br^{2})
\\\nonumber
&\times&(18-\eta+br^{2}(7\eta-6))+2a^{2}cb^{2}d(1+br^{2})^{6}(-33-34\eta+br^{2}
\\\nonumber
&\times&(24-38\eta+br^{2}(16\eta-3)))-2a^{4}cbd(1+br^{2})^{3}
(6+8\eta+br^{2}
(-9
\\\nonumber
&+&20\eta+br^{2}(15+28\eta)))+a^{5}(1+br^{2})^{2}(8c^{2}b^{2}
r^{2}\eta+d^{2}(3+4\eta+br^{2}
\\\nonumber
&\times&(12\eta-3+br^{2}(3+8\eta))))+a^{3}b(1+br^{2})^{4}
(4c^{2}b(3+4\eta+br^{2}(-6
\\\nonumber
&+&8\eta+br^{2}(9+16\eta)))-d^{2}(1+br^{2})(-3(7+6\eta)+br^{2}
(-19\eta+27
\\\nonumber
&+&br^{2}(23\eta-24))))\big]\bigg]\bigg[-40cb^{3}d(br^{2}-5)
(1+br^{2})^{10}\eta+a^{7}(2\eta-3)
\\\nonumber
&\times&d^{2}r^{2}-4a^{6}cbdr^{2}(1+br^{2})(2\eta-3)+10ab^{2}(1+br^{2})
^{8}(d^{2}(br^{2}-7)
\\\nonumber
&\times&(1+br^{2})\eta-8c^{2}b(br^{2}-1)(2\eta-3))+40a^{2}cdb^{2}d(1+br^{2})^{6}(-2\eta
\\\nonumber
&+&6+b^{2}r^{4}(11\eta-6)-br^{2}\eta)-4a^{4}cbd((1+br^{2})^{3}(5(-3+2\eta)+br^{2}
\\\nonumber
&\times&(6+br^{2}(44\eta-51)-14\eta))-2a^{3}b(1+br^{2})^{4}(-2c^{2}b((-2+17br^{2})
\\\nonumber
&\times&br^{2}+5)(-3+2\eta)+5d^{2}(1+br^{2})(6+br^{2}(-7\eta+br^{2}(-6+17\eta))))
\\\nonumber
&+&a^{5}(1+br^{2})^{2}(4(2\eta-3)c^{2}b^{2}r^{2}+d^{2}(5(2\eta-3)+br^{2}(6-24\eta+br^{2}
\\\nonumber
&\times& (74\eta-51)))))\bigg]^{-1}.
\end{eqnarray}

\section*{Appendix D}
\renewcommand{\theequation}{D\arabic{equation}}
\setcounter{equation}{0}

The radial and tangential components of adiabatic index are given by
\begin{eqnarray}\nonumber
\Gamma_{r}&=&\bigg[-6(1+br^{2})^{3}(2bd(1+br^{2})^{4}
-2acb(1+br^{2})(3br^{2}-1)+a^{2}d
\\\nonumber
&\times&(5br^{2}-1))(2\eta-1)(a^{7}d^{2}r^{2}(3-2\eta)
+4a^{6}cbdr^{2}(2\eta-3)
\\\nonumber
&\times&(1+br^{2})+8cb^{3}d(1+br^{2})^{10}(-6-13\eta
+br^{2}(17\eta-6))+8a^{2}
\\\nonumber
&\times&cb^{2}d(1+br^{2})^{6}(-9-32\eta+br^{2}(-18+41
\eta+br^{2}(23\eta-9)))
\\\nonumber
&-&2ab^{2}(1+br^{2})^{8}(d^{2}(1+br^{2})(-6-23\eta+br^{2}
(17\eta-6))+8c^{2}b
\\\nonumber
&\times&(-3-14\eta+br^{2}(26\eta-3)))-4a^{4}cbd(1+br^{2})^{3}
(3+14\eta+br^{2}
\\\nonumber
&\times&(24-46\eta+br^{2}(9+40\eta)))+2a^{3}b(1+br^{2})^{4}
(d^{2}(1+br^{2})(36\eta
\\\nonumber
&+&12+br^{2}(12-59\eta+25br^{2}\eta))+2c^{2}b(3+14\eta+br^{2}
(18-44\eta
\\\nonumber
&+&br^{2}(15+38\eta))))+a^{5}(1+br^{2})^{2}(4c^{2}b^{2}r^{2}
(3-2\eta)+d^{2}(3+14\eta
\\\nonumber
&+&br^{2}(30-48\eta+br^{2}(46\eta-9)))))\bigg]\bigg[(a^{4}dr^{2}
(2\eta-3)-2a^{3}cbr^{2}
\\\nonumber
&(&1+br^{2})(2\eta-3)+3a^{2}d(1+br^{2})^{3}(-1-2\eta+5br^{2}
(2\eta-1))
\\\nonumber
&+&2bd(1+br^{2})^{6}(-3(2+\eta)+br^{2}(17\eta-6))-2acb(1+br^{2})
^{4}(-3
\\\nonumber
&-&6\eta+br^{2}(26\eta-3)))(a^{7}d^{2}r^{2}(3-2\eta)+40cb^{3}
d(1+br^{2})^{10}\eta
\\\nonumber
&\times&(br^{2}-5)+4a^{6}cbdr^{2}(1+br^{2})(2\eta-3)+10ab^{2}
(1+br^{2})^{8}
\\\nonumber
&\times&(d^{2}(7+br^{2}(6-br^{2}))\eta+8c^{2}b(br^{2}-1)
(2\eta-3))-40a^{2}cb^{2}d
\\\nonumber
&\times&(1+br^{2})^{6}(6-2\eta-br^{2}\eta+b^{2}r^{4}(11\eta-6))
+4a^{4}cbd(1+br^{2})^{3}
\\\nonumber
&\times&(5(2\eta-3)+br^{2}(6-14\eta+br^{2}(44\eta-51)))+2a^{3}
b(1+br^{2})^{4}
\\\nonumber
&\times&(-2c^{2}b(5+br^{2}(17br^{2}-2))(2\eta-3)+5d^{2}(1+br^{2})
(6+br^{2}
\\\nonumber
&\times&(-7\eta+br^{2}(17\eta-6))))-a^{5}(1+br^{2})^{2}(4c^{2}
b^{2}r^{2}(2\eta-3)
\\\nonumber
&+&d^{2}(5(2\eta-3)+br^{2}(6-24\eta+br^{2}(74\eta-51)))))\bigg]^{-1},
\\\nonumber
\Gamma_{t}&=&\bigg[6(a^{2}d-2acb(1+br^{2})-2bd(1+br^{2})^{3})
(a^{2}r^{2}-2(br^{2}-1)
\\\nonumber
&\times&(1+br^{2})^{3})(2\eta-1)(2a^{7}d^{2}r^{2}\eta-8a^{6}cbdr^{2}
(1+br^{2})\eta-4cd
\\\nonumber
&\times&b^{3}(1+br^{2})^{10}(12+\eta+br^{2}(7\eta-6))+ab^{2}
(1+br^{2})^{8}(-8c^{2}b
\\\nonumber
&\times&(-6-8\eta+br^{2}(9+2\eta))+d^{2}(1+br^{2})(18-\eta+br^{2}
(7\eta-6)))
\\\nonumber
&\times&2a^{2}cb^{2}d(1+br^{2})^{6}(-33-34\eta+br^{2}(24-38\eta
+br^{2}(16\eta-3)))
\\\nonumber
&-&2a^{4}cbd(1+br^{2})^{3}(6+8\eta+br^{2}(-9+20\eta+br^{2}(15
+28\eta)))+a^{5}
\\\nonumber
&\times&(1+br^{2})^{2}(8c^{2}b^{2}r^{2}\eta+d^{2}(3+4\eta+br^{2}
(-3+12\eta+br^{2}(3+8\eta))))
\\\nonumber
&+&a^{3}b(1+br^{2})^{4}(4c^{2}b(3+4\eta+br^{2}(-6+8\eta+br^{2}(9
+16\eta)))-d^{2}
\\\nonumber
&\times&(1+br^{2})(-3(7+6\eta)+br^{2}(27-19\eta+br^{2}(23\eta-24)
))))\bigg]
\\\nonumber
&\times&\bigg[(4a^{4}dr^{2}\eta-8a^{3}cbr^{2}(1+br^{2})\eta+3a^{2}
d(1+br^{2})^{3}(1+2\eta+br^{2}
\\\nonumber
&\times&(2\eta-1))+2acb(1+br^{2})^{4}(-3-6\eta+br^{2}(9+2\eta))
+2(1+br^{2})^{6}
\\\nonumber
&\times&bd(3(2+\eta)+br^{2}(7\eta-6)))(-40cb^{3}d(br^{2}-5)(1+br^{2})
^{10}\eta
\\\nonumber
&+&a^{7}d^{2}r^{2}(2\eta-3)-4a^{6}cbdr^{2}(1+br^{2})(2\eta-3)+10ab^{2}
(1+br^{2})^{8}
\\\nonumber
&\times&(d^{2}(br^{2}-7)(1+br^{2})\eta-8c^{2}b(br^{2}-1)(2\eta-3)
+40a^{2}cb^{2}d
\\\nonumber
&\times&(1+br^{2})^{6}(6-2\eta-br^{2}\eta+b^{2}r^{4}(11\eta-6))
-4a^{4}cdb(1+br^{2})^{3}
\\\nonumber
&\times&(5(2\eta-3)+br^{2}(6-14\eta+br^{2}(44\eta-51)))-2a^{3}
b(1+br^{2})^{4}(-2
\\\nonumber
&\times&c^{2}b(5+br^{2}(17br^{2}-2))(2\eta-3)+5d^{2}(1+br^{2})
(6+br^{2}(-7\eta
\\\nonumber
&+&br^{2}(17\eta-6))))+a^{5}(1+br^{2})^{2}(4c^{2}b^{2}r^{2}
(2\eta-3)+d^{2}(5(2\eta-3)
\\\nonumber
&+&br^{2}(6-24\eta+br^{2}(74\eta-51))))))\big]^{-1}.
\end{eqnarray}
\textbf{Data Availability Statement:} No new data was created or
analyzed in this study.

\end{document}